\documentclass[reprint,aps,a4paper,pra,twocolumn]{revtex4-1}

\usepackage{graphicx}
\usepackage{float}
\usepackage{subfigure}
\usepackage{dcolumn}
\usepackage{gensymb}
\usepackage{bm}
\usepackage{epstopdf}
\usepackage{amsmath}

\usepackage{color}
\usepackage[usenames,dvipsnames,svgnames,table]{xcolor}

\usepackage{hyperref}
\hypersetup{
     colorlinks   = true,
     citecolor    = blue,
     linkcolor    = blue
}

\begin{document}
\title{All-thermal switching of amorphous Gd-Fe alloys: analysis of structural properties and magnetization dynamics}

\author{Raghuveer Chimata,$^1$  Leyla Isaeva,$^1$ Krisztina K\'adas,$^{1,3}$ Anders Bergman,$^1$ Biplab Sanyal,$^1$ Johan H. Mentink,$^2$ Mikhail I. Katsnelson,$^2$ Theo Rasing,$^2$ Andrei Kirilyuk,$^2$ Alexey Kimel,$^2$ Olle Eriksson,$^1$ and Manuel Pereiro$^1$}
\affiliation{$^1$Department of Physics and Astronomy, Uppsala University, Box 516,
 751\,20 Uppsala, Sweden}
\affiliation{$^2$Radboud University Nijmegen, Institute of Molecules and Materials, Heyendaalseweg 135, 6525 AJ Nijmegen, The Netherlands}
\affiliation{$^3$Institute of Solid State Physics and Optics, Wigner Research Centre for Physics, Hungarian Academy of Sciences, 1525 Budapest, P.O.B. 49, Hungary}

\date{\today }

\begin{abstract}
In recent years, there has been an intense interest in understanding the microscopic mechanism of thermally induced magnetization switching driven by a femtosecond laser pulse. Most of the effort has been dedicated to periodic crystalline structures while the amorphous counterparts have been less studied. By using a multiscale approach, i.e. first-principles density functional theory combined with atomistic spin dynamics, we report here on the very intricate structural and magnetic nature of amorphous Gd-Fe alloys for a wide range of Gd and Fe atomic concentrations at the nanoscale level. Both structural and dynamical properties of Gd-Fe alloys reported in this work are in good agreement with previous experiments. We calculated the dynamic behavior of homogeneous and inhomogeneous amorphous Gd-Fe alloys and their response under the influence of a femtosecond laser pulse. In the homogeneous sample, the Fe sublattice switches its magnetization before the Gd one. However, the temporal sequence of the switching of the two sublattices is reversed in the inhomogeneous sample. We propose a possible explanation based on a mechanism driven by a combination of the Dzyaloshiskii-Moriya interaction and exchange frustration, modeled by an antiferromagnetic second-neighbour exchange interaction between Gd atoms in the Gd-rich region. We also report on the influence of laser fluence and damping effects in the all-thermal switching. 
\end{abstract}
\maketitle

\section{Introduction}

Switching the sublattice magnetization directions of amorphous Gd-Fe alloys~\cite{radu} 
(doped with small amounts of Co) by intense femtosecond laser pulses has generated significant interest 
both experimentally and theoretically. Amorphous Gd-Fe alloys are ferrimagnetic, with a strong antiferromagnetic 
(AFM) coupling between the rare-earth and transition metal moments, a coupling which has its explanation 
in the hybridization of the {\it 5d} and {\it 3d} orbitals of the constituting elements~\cite{spunzar}. In Ref.~[\onlinecite{radu}], 
it was found that an optical excitation caused the net magnetization of both the Gd and Fe sublattices 
to rapidly collapse. However, the time scales of the dynamics of the two sublattices were found to be quite 
different: the net magnetic moment of the Fe sublattice was found to vanish after 0.4 ps and then for 
a short period of time, up to 2 ps, be parallel to the Gd moment. The Gd sublattice, which initially is 
antiferromagnetic to Fe, vanished after 2 ps, after which it reversed to have its magnetization opposite 
to that of Fe, hence completing the reversal process (see Fig.~3 of Ref.~[\onlinecite{radu}]). The interest of 
these results obviously have great potential for technological applications, since they open up for possibilities 
to store information in a magnetic medium without applying an external magnetic or electric field. 
In fact, the experimental results reported in Ref.~[\onlinecite{radu}] follow intense investigations of magnetization dynamics, 
which started in the mid '90s~\cite{beaurepaire,vahaplar, malinowski,bigot,stamm,boeglin}.

Different theoretical models have attempted to explain these results. For instance, in Ref.~[\onlinecite{mentink}]
it was argued that two time and temperature domains were relevant, where the spin-relaxation was driven first 
by a relativistic contribution whereas subsequently relaxation was argued to be governed by an exchange origin. 
A different explanation was provided in Ref.~[\onlinecite{ostater}] where the coupling between Gd and Fe dominated 
magnon modes were identified as the most important aspect of the complex switching behavior of the Gd-Fe system. 
It should also be mentioned that in the experimental investigation of Ref.~[\onlinecite{graves}], it was speculated 
that angular momentum was transferred between the different sublattices via spin-currents, and
this was identified as the most important aspect of the magnetization dynamics of amorphous Gd-Fe alloys.

Although the main experimental findings of Ref.~[\onlinecite{radu}] have been repeated in subsequent experiments, 
there are details in a more recent work that have so far not been addressed satisfactorily by theory. For instance, 
in Ref.~[\onlinecite{graves}] several hitherto unexplained experimental facts were reported. Moreover, in the 
samples measured by Graves et al.~\cite{graves} concentration profiles were detected, with Gd rich/Fe poor regions 
and Gd poor/Fe rich regions, in the same sample. Surprisingly, it was found that for the Gd rich regions, the Gd 
moment has a different dynamical response compared to that of the Gd poor regions. This amounted to situations 
where in the Gd rich regions, the Gd moment reversed before the Fe moment, in contrast to the result reported 
for the average magnetization of Gd or Fe sublattices, reported in Ref.~[\onlinecite{radu}]. Hence, in amorphous Gd-Fe alloys, 
it seems that sometimes the Fe moment reverses before the Gd moment, and sometimes Gd switches before Fe, depending on the concentration of Gd and Fe in local regions of the sample.

None of the theories presented so far has addressed the role of the amorphous structure and the chemical inhomogeneity of the Gd-Fe system and how this influences the ultrafast switching behavior. In this work, we present a multi-scale approach to address this problem, where we coupled first principles electronics structure theory to atomistic spin dynamics simulations \cite{skubic}. After introducing our methodology, we substantiate our approach by comparing equilibrium magnetization curves with experiment for a wide range of concentrations, elucidating the crucial role of the amorphous atomic arrangement on the magnetism. Subsequently, we demonstrate that the results for homogenous samples are in agreement with previous theoretical analysis reported in Ref.~\cite{mentink}. Finally we turn to inhomogenous samples and demonstrate that the switching is crucially affected by the chemical inhomogeneity and the non-collinearity of the spins in the rare-earth sublattice. In the appendix, the methods are explained in more detail and an analysis of the role of the damping is also provided. 

\section{Method}

\subsection{Details of the simulation of structural properties}

First-principles spin polarized calculations were performed by means of the density functional 
theory ~\cite{hohenberg64,kohn65} and projector augmented wave ~\cite{blochl94,kresse99} method 
as implemented in the Vienna ab initio simulation package (VASP) ~\cite{vasp1,vasp2,vasp3}.
The exchange-correlation potential was treated using the generalized gradient approximation with 
the Perdew, Burke, and Ernzerhof functional ~\cite{pbe}, including the valence states 
$5s^{2} 5p^{6} 5d^{1} 6s^{2} 4f^{7}$ for Gd and $3d^{7} 4s^{1}$ for Fe. The LDA+U method ~\cite{dudarev98} 
was applied to Gd with {\it U$_{eff}$}=7 eV and {\it J}=1 eV.

The amorphous structures were generated by means of the stochastic quenching (SQ) 
method~\cite{holmstrom2009,holmstrom2010a}, as described in Ref.~[\onlinecite{kadas2012}].
This method is based on the single-random-valley approximation in vibration-transit (VT) 
theory ~\cite{wallace97,lorenzi2007}. The SQ method was demonstrated to provide
reliable atomic coordinates of amorphous materials ~\cite{kadas2012}.
In the initial structures, 200 atoms were both spatially and chemically randomly distributed in a 
cubic unit cell with periodic boundary conditions and a density of $\varrho$=7.87 g/cm$^3$ for Gd$_{0.24}$Fe$_{0.76}$,
$\varrho$=7.88 g/cm$^3$ for Gd$_{0.50}$Fe$_{0.50}$, and $\varrho$=7.89 g/cm$^3$ for Gd$_{0.76}$Fe$_{0.24}$.
The atomic positions were then relaxed until the force on every atom was negligible.
The calculations were performed using the $\Gamma$ point.

The kinetic energy cutoff of 550 eV together with Methfessel-Paxton band smearing ~\cite{methfessel} 
of $\sigma$ = 0.2 eV were used for electronic structure calculations.   
The atomic charges were determined from Bader analysis ~\cite{bader1,bader2,bader3}.

\subsection{Details of the atomistic spin-dynamics simulations }

In our simulations, we combined the two-temperature (2T) model~\cite{2t} with the atomistic spin dynamics (ASD) in the UppASD
code~\cite{skubic}  using the Landau-Lifshitz-Gilbert (LLG) equation. Model exchange parameters were used for all simulations. At a finite 
temperature, the temporal evolution of individual atomic moments in an effective field is governed by Langevin dynamics,

\begin{equation}
\frac{d{\bf m}_{i}}{dt} = -\gamma{\bf m}_{i}\times[{\bf B}_{i}+{\bf b}_{i}]-
\gamma\frac{\alpha}{{\textit m}}
{\bf m}_{i}\times ({\bf m}_{i}\times [{\bf B}_{i}+{\bf b}_{i}]),
\end{equation}
where $\gamma$ is the gyromagnetic ratio, $\alpha$ represents the dimensionless phenomenological 
Gilbert damping constant and ${\bf{m}}_i$ stands for an individual atomic moment on site $i$. The ``effective" magnetic field is represented by $\bf{B}_i$ while ${\bf b}_{i}$ is a time evolved stochastic magnetic 
field, which depends on the electron temperature from the 2T model. After applying a femtosecond 
laser pulse on the samples, the electron temperature increase from the initial temperature $T_0$ to a peak 
temperature in less than 50 fs. Then, the electron temperature slowly cools 
down in about $5\cdot10^3$ fs since the heat of the electron system is transferred to the phonon bath via electron-phonon interactions~\cite{2t}. 
With this method, details of all thermal switching are investigated in detail, and reported upon below.

\section{Results and discussion}

\subsection{Ab-initio theory and structural properties}\label{structural}

In this section we provide structural information of Gd$_x$Fe$_{1-x}$ ($x = 0.24, 0.50, 0.76$) 
magnetic alloys based on {\it ab initio} theory. We note here that the first-principles calculations 
result in a metallic character of these amorphous materials, in agreement with experimental 
observations. The electronic properties of Gd-Fe systems are described in more 
details in the Appendix~\ref{appendixa}. 

\begin{figure}[htb]
\begin{center}
\includegraphics[angle=0,scale=0.47]{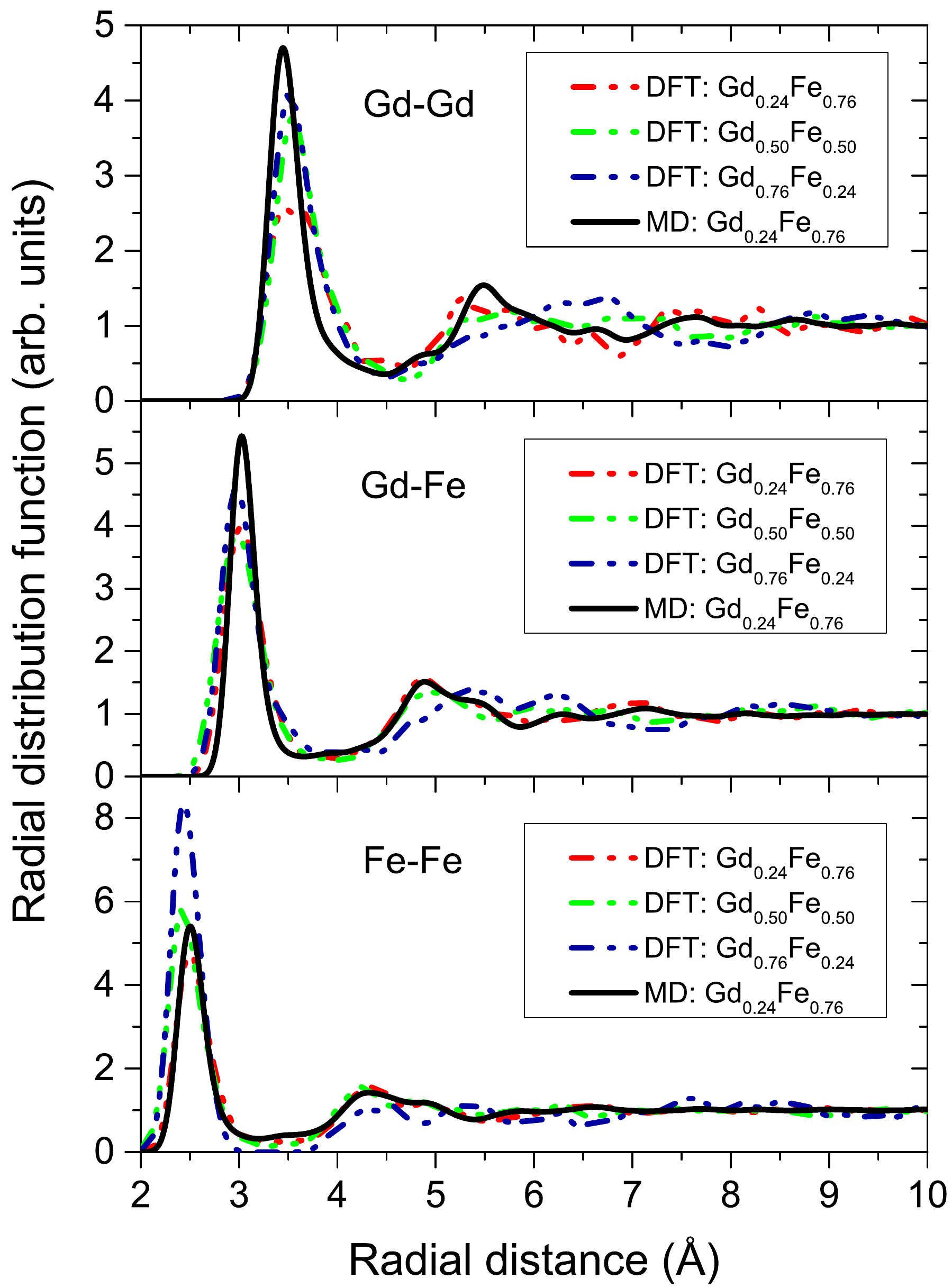}
\caption{\label{figure1} (Color online) Radial distribution function in Gd$_{x}$Fe$_{1-x}$ 
with three different stoichiometries. The dashed lines show data calculated with DFT, while the black solid line represents data provided by molecular dynamics calculations.}
\end{center}
\end{figure}

The local atomic environment in amorphous Gd$_x$Fe$_{1-x}$ can be analyzed 
by using radial distribution functions (RDF) calculated for different atomic pairs (see Fig.~\ref{figure1}). 
From RDFs of the SQ-generated structures, we find short-range order up to 8 \AA~ for Gd-Gd and Gd-Fe, 
and up to 6 \AA~ for Fe-Fe atomic pairs. Gd-Gd, Gd-Fe and Fe-Fe bond lengths, 
extracted from RDFs, are shown in Table \ref{table1} along with the bond lengths in selected 
reference systems.  The theoretical Gd-Gd bond length in amorphous Gd$_{x}$Fe$_{1-x}$ is found to be shorter, and therefore the bonds are stronger 
than in hexagonal close-packed Gd. At the same time, the Gd-Gd bond length is longer than in 
crystalline compounds consisting of Fe and Gd, such as cubic GdFe$_{2}$, and trigonal GdFe$_{3}$, 
suggesting a weaker bonding in the amorphous matrix.
We find a favorable agreement between theoretical Gd-Gd bond distance in the Fe-rich ($x = 0.24$)
and equiatomic ($x = 0.50$) amorphous systems compared to the experimental values for melt-quenched 
amorphous Gd$_{0.22}$Fe$_{0.78}$ and Gd$_{0.56}$Fe$_{0.44}$, respectively (see Table \ref{table1}).
The theoretical Gd-Fe bond length in amorphous Gd$_x$Fe$_{1-x}$ is very close to that 
in crystalline GdFe$_{2}$ and GdFe$_{3}$. We find the bond distance between Fe atoms in Gd$_{x}$Fe$_{1-x}$ 
to be shorter than in bcc Fe and cubic GdFe$_{2}$, but at the same time larger than 
in trigonal GdFe$_{3}$. 
We also find a remarkable agreement between theoretical bond lengths between different pairs of atoms, such as
Gd-Gd, Gd-Fe and Fe-Fe, in Gd$_x$Fe$_{1-x}$ ($x = 0.24, 0.50$) obtained by SQ simulations and the experimental 
ones for quench-melted Gd$_{x}$Fe$_{1-x}$ ($x = 0.22, 0.56$) with similar stoichiometry. This illustrates the 
efficiency and accuracy of the SQ method to describe the structural properties of amorphous materials.

\begin{table}[ht]
\caption{ { Bond lengths (\AA) in amorphous Gd$_x$Fe$_{1-x}$} system. The bond distances in selected crystalline
and amorphous systems are listed for comparison.} 
\centering 
\begin{tabular}{l c c c c } 
\hline\hline 
System     &~~~~~& Gd-Gd &Gd-Fe &  Fe-Fe                        \\
\hline
Gd$_{0.24}$Fe$_{0.76}$ &&      3.47 &          3.02 &          2.50          \\ %
Gd$_{0.50}$Fe$_{0.50}$& &      3.55 &         2.98 &         2.43         \\ 
Gd$_{0.76}$Fe$_{0.24}$ &&      3.52 &          2.97  &          2.44        \\ 
\hline
hcp Gd  \cite{spedding}&& 3.57 &-&-\\
bcc Fe \cite{basinski} &&-&-&2.54   \\
GdFe$_{3}$ \cite{smith}&& 3.21   & 2.97   & 2.37\\
GdFe$_{2}$ \cite{endter}&& 3.22   & 3.08    & 2.63    \\
am-Gd$_{0.22}$Fe$_{0.78}$ \cite{petkov}&& 3.47   & 3.11  & 2.57   \\
am-Gd$_{0.56}$Fe$_{0.44}$ \cite{petkov}&& 3.54   &  2.95   & 2.51  \\
\hline
\end{tabular}\label{table1} 
\end{table}

Next, we analyze the local environment in amorphous Gd$_{x}$Fe$_{1-x}$, as it is represented by the average 
coordination numbers (Table \ref{table2}). With increasing Gd concentration, the Gd-Gd average coordination 
number, as expected, increases from 4.6 for $x = 0.24$ to 10.9 for $x = 0.76$. Similarly, the Gd-Fe 
coordination number increases with the number of Gd atoms. In case of the Gd-Fe atomic 
pair, the coordination number is almost five times smaller in the Fe-rich amorphous matrix compared to 
the Gd-rich one. Also, while the Fe concentration decreases within the series (from  $x = 0.24$ to $x = 0.76$), 
so does the Fe-Fe coordination number. However, the total coordination numbers on both Gd and Fe coordination shells
overall change similarly, {\it i.e.} decrease through the series. The reduction in the total coordination number both for Gd and Fe atoms can be referred to the change from a more close packed structure (Fe-rich system) to a more open one (Gd-rich system). 

\begin{table}[ht]
\caption{ {Average coordination numbers for amorphous Gd$_x$Fe$_{1-x}$} system. Coordination numbers
in reference systems are listed for comparison in the lower part of the table. In GdFe$_3$ 
the coordination for Gd and Fe atoms with different site symmetries  
is different. Therefore, we show coordination for all inequivalent positions (specified in parentheses).} 
\centering 
\begin{tabular}{l c c c c } 
\hline\hline 
        System                   &  Gd-Gd         & Gd-Fe          & Fe-Gd          & Fe-Fe       \\ [0.5ex] 
\hline 
Gd$_{0.24}$Fe$_{0.76}$     &  4.6           & 11.0           & 3.5            & 7.6          \\ 
Gd$_{0.50}$Fe$_{0.50}$     &  9.1           & 5.7            & 5.7            & 4.5          \\
Gd$_{0.76}$Fe$_{0.24}$     & 10.9           & 2.4            & 7.5            & 2.0          \\
\hline
GdFe$_{2}$ ~\cite{inoue}   & 4              & 12             &  6             &  6           \\
GdFe$_{3}$ ~\cite{smith}   & (3{\it a}) 2     & (3{\it a}) 6-12  & (3{\it b})  6    & (3{\it b})  6   \\
                           & (6{\it c}) 1-3   & (6{\it c}) 3-6   & (6{\it c})  3    & (6{\it c})  3   \\
                           &                    &                    & (18{\it h}) 1-2  & (18{\it h}) 1-2  \\
am-Gd$_{0.22}$Fe$_{0.78}$ ~\cite{petkov} & 3.0            & 8.8            & 2.5            & 7.9            \\
am-Gd$_{0.56}$Fe$_{0.44}$ ~\cite{petkov} & 7.5            & 3.2            & 4.2            & 3.0            \\
\hline
\end{tabular}\label{table2} 
\end{table}

\subsection{Generation of amorphous samples using molecular dynamics}

As commented in Sec.~\ref{structural}, we optimised the structures of Gd$_x$Fe$_{1-x}$ (x=0.24, 0.50, 0.75) magnetic alloys by means of {\it ab initio} methods considering a supercell of 200 atoms, but the lack of crystal periodicity in amorphous structures led us to consider even bigger supercells just to be sure that the results are reliable and the physics of the amorphous structure was fully captured. The size of the new supercells is beyond the limits of the present state of the art of the computational resources using a DFT methodology. In order to deal with bigger supercells, we employed a molecular dynamics approach, so that the dynamics of atomistic Fe and Gd spins shown in upcoming sections have been performed using as input parameters the structural data provided by the molecular dynamics method. Consequently, we adapted a two step procedure. Initially, we constructed a cubic unit cell (1600 atoms) with Gd and Fe atoms using a dense-random-packing-of-hard-spheres (DRPHS) model and using as input the lattice parameter ($\sim 29.5$~\AA) provided by geometry optimization in VASP calculations for a cell of 200 atoms~\cite{kenneth}. In the second step, the forces on atoms were minimized by using an adapting Morse pair potential in Large-scale Atomic/Molecular Massively Parallel Simulator (LAMMPS) code~\cite{brown}. During the optimization process, experimental bond lengths between the species and potential depths were taken from Ref.~\cite{chen}. The relaxed molecular dynamics (MD) amorphous samples were analyzed via their RDF main peak positions and nearest-neighbour distributions. We illustrate in Fig.~\ref{figure1} the radial distribution calculated by DFT and MD simulations for Gd$_{0.24}$Fe$_{0.76}$. It may be observed that the agreement is rather good between the two sets of  theoretical values.

\subsection{Magnetization-dynamics and all-thermal switching}

\subsubsection{Curie and compensation temperatures}
 
\begin{figure}[ht]
\begin{center}
\includegraphics[angle=0,scale=0.33]{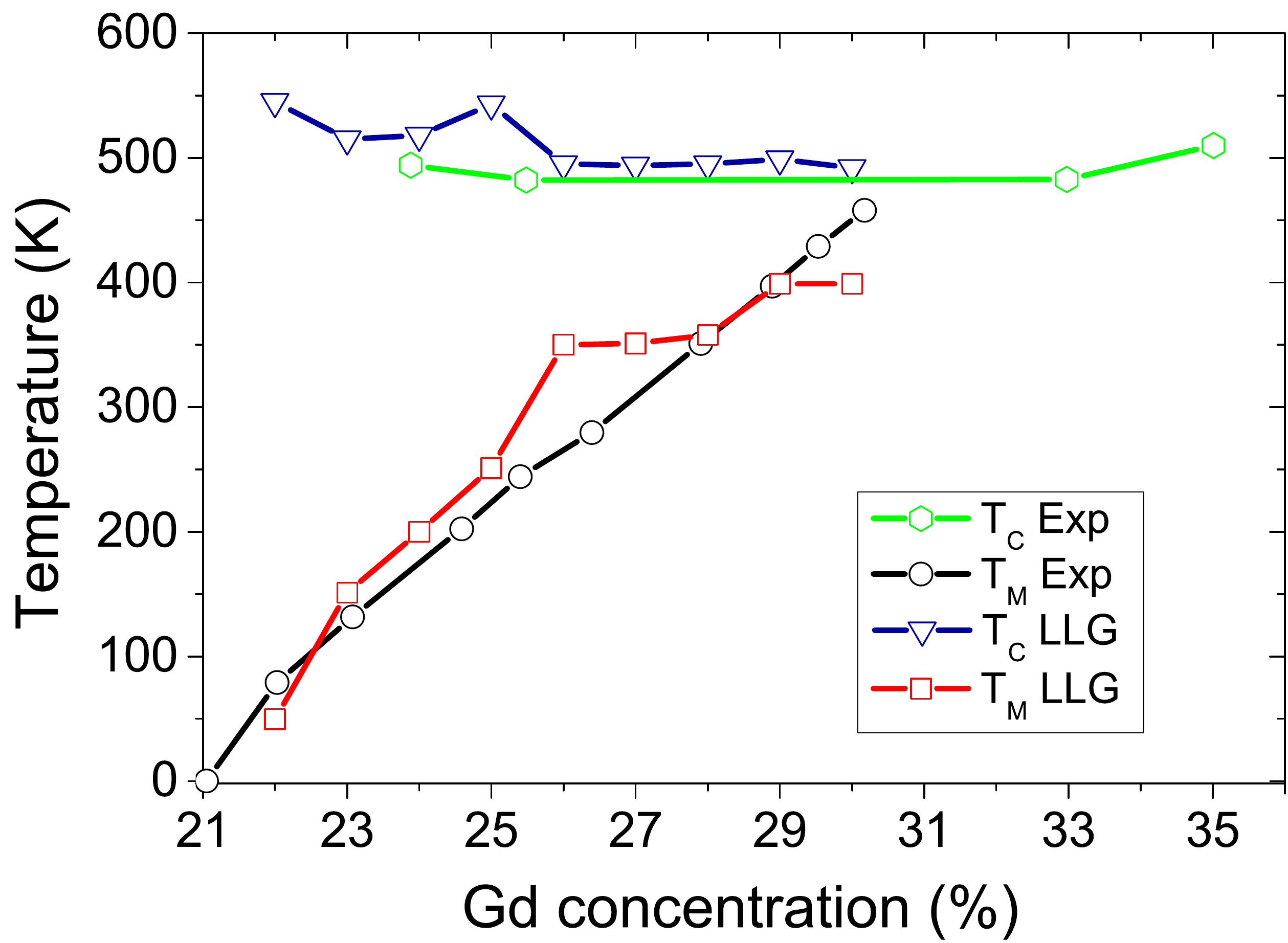} 

\end{center}

\caption{(Color online) Curie temperature ({\it T$_C$}) and magnetic compensation temperature ({\it T$_M$}) for amorphous Gd-Fe alloys, for different concentrations of Gd. The experimental data has been taken from Ref.~[\onlinecite{mimura}].}\label{figure2}
\end{figure}

To further extend the applicability of our methodology, we performed ASD simulations using the UppASD method on a cell containing 1600 atomic spins with periodic boundary conditions. We take a ferrimagnetic amorphous GdFe model system based on the structural parameters provided by both molecular dynamics and DFT calculations. The microscopic model exchange parameters were taken from Ref.~\cite{ostler1}. We used the bulk exchange values for neighbouring TM and RE ions ({\it J$_{Fe-Fe}$}=0.8 mRy, {\it J$_{Gd-Gd}$}=0.15 mRy) because they provide the correct Curie temperature for the respective sublattices. The value of the intersublattice exchange coupling ({\it J$_{Gd-Fe}$} =-0.25 mRy) was chosen to fit the temperature dependence of the saturation magnetisation of both Fe and Gd sublattices with results of x-ray magnetic circular dichroism measurements of static magnetisation reported in Ref.~\cite{ostler1}. For the magnetic moments we take the bulk values, i. e.  7.6~$\mu_B$  and 2.1~$\mu_B$  for Gd and Fe, respectively. No external field was applied in the simulations and the anisotropy of the Gd and Fe sublattices was neglected in the hamiltonian. The model exchange parameters are capable of reproducing the key static magnetic properties, especially the Curie temperatures and magnetic compensation points. To illustrate this fact we show in Fig.~\ref{figure2} both the calculated compensation and Curie temperatures for different Gd concentrations, in the range 20 at.$\%$ to 30 at.$\%$. The Curie temperatures have been calculated using a finite size scaling analysis as described in Ref.~[\onlinecite{binder}] (see Appendix~\ref{bindercumulants}). We found a very good agreement with the reported experimental results. Moreover, we observe a general trend for the T$_\mathrm{C}$ to decrease as the Gd concentration increases. We attribute this magnetic softening to the addition of more Gd-Gd nearest neighbours, which have a smaller exchange coupling as compared to the Fe-Fe interaction. But still, since we have an amorphous structure and a supercell with limited number of atoms, this condition is not fulfilled for every Gd concentration and this is why at some concentrations, the T$_\mathrm{c}$ can still increase slightly, as is the case for a concentration of Gd of about 25 at.\%. In the thermodynamic limit, these smaller fluctuations of the {\it T$_C$} are expected to vanish. It may be seen from Fig.~\ref{figure2} that the simulations reproduce with good accuracy the measured compensation temperatures as well. Both the measured trend and the absolute values of the compensation temperature of these alloys are reproduced by theory, where the most noticeable feature is the increase of the compensation temperature with increasing Gd concentration. The reason for this trend is a competition of magnetic sublattices with antiparallel coupling. Too few Gd atoms result in a Gd sublattice with a net magnetization that is smaller than that of the Fe sublattice, already at low temperatures, and there is no compensation point. However, with increasing Gd concentration, the net magnetic moment of this sublattice is larger than that of the Fe sublattice, at low temperatures. Since the Fe exchange is stronger than the Gd exchange, the Gd sublattice magnetization decays faster with temperature compared to the Fe sublattice magnetization, and at the compensation temperature, they have equal size and opposite direction. Increasing the Gd concentration makes the magnetization of this sublattice stronger relative to the Fe sublattice, at low temperatures. Hence a higher temperature is needed in order to reduce the Gd moment to have the same size, albeit with opposite direction, compared to the Fe sublattice. 

It is rewarding that the agreement between theory and experiment found in Fig.~\ref{figure2} is quite good, and that the three parameters of exchange interactions used in our simulations explain the compensation temperatures of the whole range of concentrations of Fig.~\ref{figure2}. We also note that finer details of the atomic arrangement of the amorphous structure are very important in achieving the results shown in Fig.~\ref{figure2}, and this illustrates (as often is the case) that atomic arrangement (structure) and magnetic properties are intimately coupled.

\begin{figure}[ht]
\begin{center}
\includegraphics[angle=0,scale=0.32]{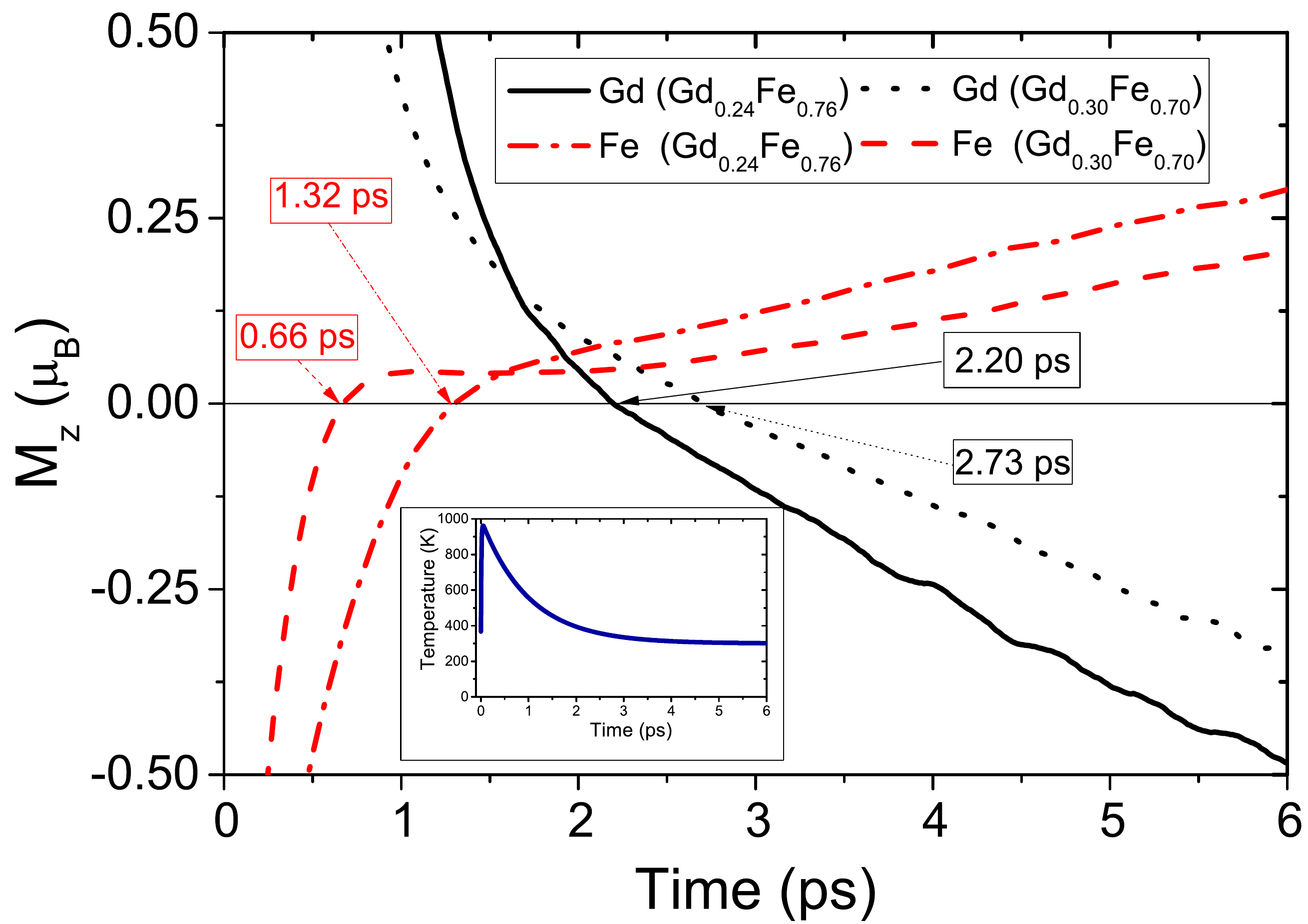}
\caption{\label{figure3} (Color online) Time evolution of the magnetization (M$_z$) for two different concentrations of amorphous Gd-Fe alloys under the influence of a thermal heat pulse.  The solid and dash-dot lines represent the sample concentration Gd$_{0.24}$Fe$_{0.76}$, while the dash and dot lines are for Gd$_{0.30}$Fe$_{0.70}$. The magnetization of the two  sublattices is plotted separately. In the inset we show a typical temperature profile induced from the laser fluence, as given by the two temperature model of Ref.~\cite{2t}.}
\end{center}
\end{figure}

\subsubsection{All-thermal switching of homogenous samples}

In this section, we discuss the dynamic behavior of homogenous amorphous alloys, using the exchange parameters discussed above, and the magnetic response to a femtosecond laser pulse. Only temperature effects from the laser pulse were considered, where we adopted a two-temperature model, as described in the Appendix~\ref{2tmodel}. All simulations started with the spin system at room temperature, from which the heat pulse increased the temperature of the spin system in a very short period of time (50 fs) to a maximum value, T$_{max}$, after which the sample cooled down again. We observed sublattice switching for a wide range of concentrations, i.e. 21 at.\% to 30 at.\% Gd. As an example we show in Fig.~\ref{figure3} the switching behavior of the Gd$_{0.24}$Fe$_{0.76}$ and the Gd$_{0.30}$Fe$_{0.70}$ alloys. We find for both concentrations that initially both Gd and Fe sublattices demagnetize fast, and that the Fe sublattice reverses its magnetization first, so that for a short period of time both Fe and Gd sublattices have parallel magnetic moments. Figure~\ref{figure3} shows that after $\sim$3-4 ps the all-thermal switching is more or less completed, and the reversed sublattice moments relax to their new equilibrium directions, as the spin system cools down. Alloys with different concentrations have slightly different behavior, although the main features are independent on concentration.
All features of Fig.~\ref{figure3} are in agreement with the observations of Ref.~[\onlinecite{radu}] and also with the simulations and theoretical analysis described in Ref.~[\onlinecite{mentink}].

The initial temperature (300 K) of some of our simulations shown in Fig.~\ref{figure3} is above, or equal to, the compensation temperature (alloys with 21 at.\% to 26 at.\% Gd) while for other simulations the initial temperature is below the compensation temperature (alloys with 26 at.\% to 30 at.\% Gd). The all-thermal switching behavior was observed for all cases, irrespective whether the initial temperature was above or below a compensation temperature. The rule-of-thumb proposed for all thermal switching in Ref.~\cite{medapalli}, that it is critical to start with an initial temperature below the compensation temperature, does not seem to hold in light of the present work.

\begin{figure}[ht]
	\begin{center}
		\includegraphics[angle=0,scale=0.38]{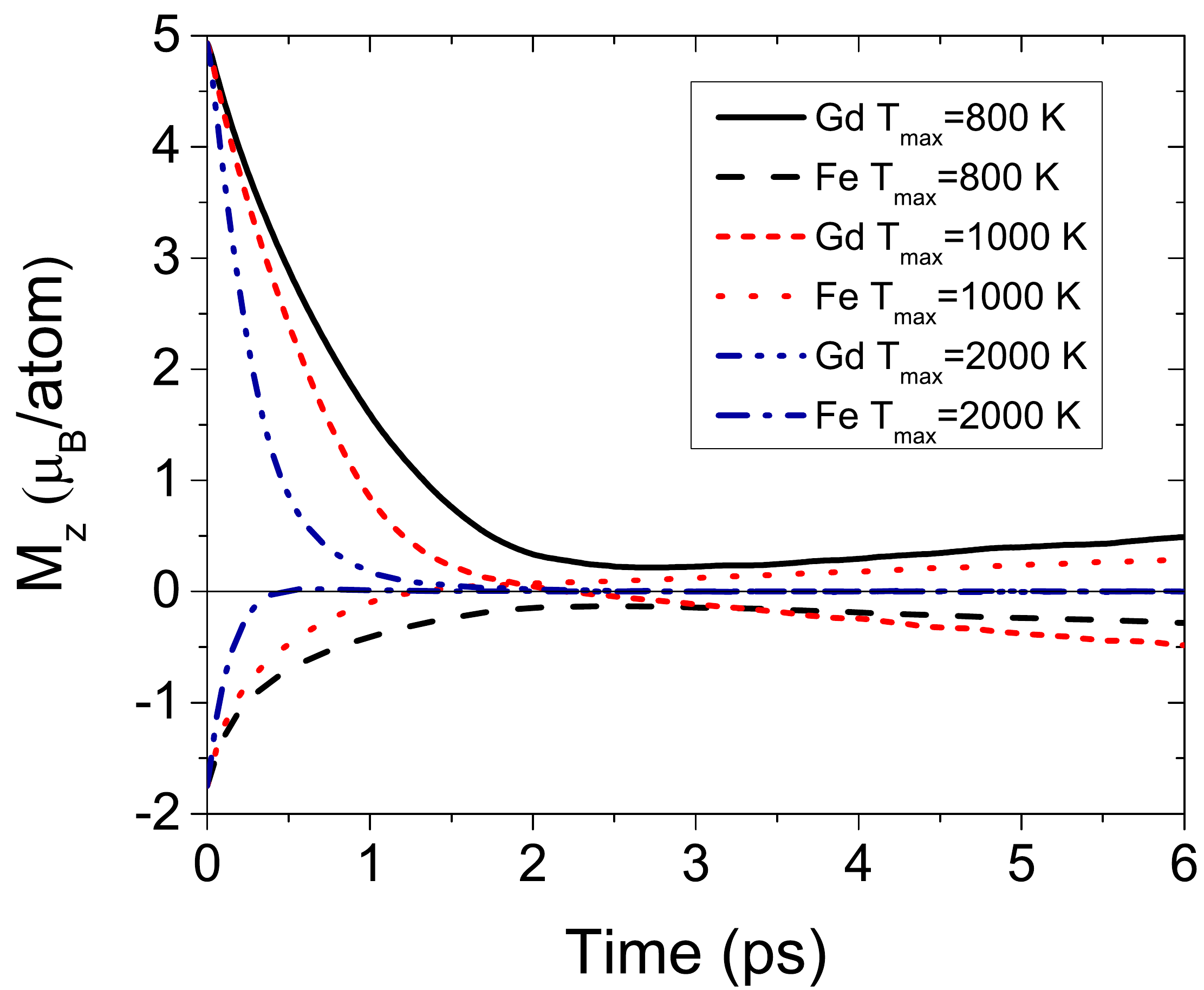}
		\caption{\label{figure4} (Color online) Time evolution of the magnetization (M$_z$) of amorphous Gd-Fe alloys (Gd$_{0.24}$Fe$_{0.76}$) for different peak temperatures caused by the laser fluence.}
	\end{center}
\end{figure}

In Fig.~\ref{figure4}, we show results of the simulated magnetization dynamics for various peak temperatures of the spin system. Note that if the peak temperature is not sufficiently high there is no reversal, the sublattices demagnetize to some fraction of their original value, and then simply return to the initial configuration, as the spin temperature is lowered. This is the case e.g. for $T_{max}$ = 800 K. If the spin temperature is too high (e.g. the simulation with $T_{max}$ = 2000 K) both sublattices simply demagnetize to a zero moment state, which is stable for sufficiently long time in order to make the remagnetization completely stochastic in terms of direction of each sublattice moment of the final configuration. However, for intermediate temperatures (e.g. $T_{max}$ = 1000 K) the all thermal switching occurs, as is also shown in Fig.~\ref{figure4}. This shows that it is essential to find the appropriate laser fluence with respect to the strength of the exchange interactions, for all-thermal switching to occur.

We end this section with a short note on the effects of the damping. We investigated the magnetization dynamics for a wide range of damping parameters, as detailed in the Appendix~\ref{damping_effect}, and found that the switching behavior reported in Fig.~\ref{figure3} was essentially very dependent on the choice of damping parameter. This shows that the strength of the intrinsic damping actually determines if the all-thermal switching can either occur or not. 

\subsubsection{All-thermal switching of inhomogeneous samples}
\label{inhomogeneous_section}

After the initial experimental work of Ref.~[\onlinecite{radu}] additional experimental data was reported, and in particular it was argued in 
Ref.~[\onlinecite{graves}] that amorphous Gd-Fe alloys may have non-uniform concentration profiles, such that some regions are richer in Gd and some 
are poorer, with the opposite trend for the Fe concentration. Interestingly, it was reported that in an all-thermal switching experiment of an amorphous Gd-Fe alloy with heterogenous concentration, the Gd magnetic moment reached zero before the Fe magnetic moment in the Gd rich regions. After a period of parallel alignment of Fe and Gd magnetic moments, the reversal completed with both Fe and Gd moments having reversed orientations with respect to their original direction. Hence, Gd rich regions exhibited similar behavior as shown here in Fig.~\ref{figure3}, albeit with the Gd sublattice reaching zero first. 

In order to investigate this experimental result and mimic the experimental samples as closely as possible, we generated simulation cells with concentration profiles, such that, some regions had enhanced (depleted) Gd (Fe) concentration with respect to the nominal concentration while some other regions had a depletion (enhancement) of Gd (Fe) concentration. We considered an amorphous alloy with average concentration Gd$_{0.24}$Fe$_{0.76}$ and the Gd rich regions had an increase of 6 at.\% Gd, while the Gd poor regions had a reduction of 6 at.\% of the Gd concentration. The Fe concentration was modified in the same way. A schematic illustration of such a heterogenous sample is shown in Fig.~\ref{figure5}, and further details on how the inhomogeneous simulation-cells were generated can be found in the Appendix~\ref{inhomogeneous}. In these simulations, we have in the initial configuration also considered different degrees of non-collinearity of the Gd moments. This is supported by our first principles calculations, presented above, that show an exchange driven non-collinear configuration as the ground state even at T=0 K. From first-principles non-collinear theory, we find that the degree of non-collinearity is more pronounced for calculations including spin-orbit (LS) coupling than without it. Moreover, the effect is more prominent for Gd sublattices (see Table~\ref{table3} and Appendix~\ref{appendixa}). For example, it is worthy to mention here that the current DFT calculations with spin-orbit coupling predict a maximal angle deviation of Gd atomic magnetic moments with respect to the quantization axis of about 35$\degree$. However, the angle deviation predicted by DFT without spin-orbit coupling is drastically reduced down to less than the half of the value with LS couping. Since the amorphous structure lacks inversion symmetry, the Dzyaloshinskii-Moriya (DM) interaction, which favours canted (non-collinear) configurations, is non-zero for both Gd and Fe sublattices. Since the spin-orbit interaction is larger for Gd, it is for this sublattice we expect a larger effect of the DM interaction. In Table~\ref{table3} we have collected the average  and maximum angles of the magnetic moments, with respect to a common z-axis, for various concentrations. It may be seen that exchange effects alone produce a certain degree of non-collinearity of the moments, where in particular the maximum deviation from collinearity is larger for Gd than for Fe. With spin-orbit effects included, the degree of non-collinearity increases, in particular for the Gd sublattice. In practice, we have included the degree of non-collinearity, from exchange as well as DM interaction, among the Gd atomic moments via a second nearest neighbour (SNN) exchange interaction (${\it J_2}$) with anti-ferromagnetic character, and as reported below, we have followed the simulated magnetization dynamics as a function of the degree of non-collinearity of Gd sublattices.

\begin{table}[ht]
\caption{ Degree of non-collinearity predicted by the current DFT calculations with and without spin-orbit (LS) coupling. The collected data represent the average (A) and maximal (M) angle deviation with respect to the z-axis of the magnetic moments on Gd and Fe atoms for amorphous Gd$_x$Fe$_{1-x}$ structures ($x=0.24, 0.50, 0.76$).} 
\centering 
\begin{tabular}{l c c c c c c c c c c c c} 
\hline\hline 
System     & &  \multicolumn{5}{c}{LS coupling}  &  & \multicolumn{5}{c}{no LS coupling}  \\
\cline{1-1}\cline{3-7}\cline{9-13}
&&\multicolumn{2}{c}{Gd} &&\multicolumn{2}{c}{Fe}& & \multicolumn{2}{c}{Gd} &&\multicolumn{2}{c}{Fe} \\
\cline{3-4}\cline{6-7} \cline{9-10}\cline{12-13}
& &A&M& &A&M& & A&M&&A&M\\
\hline
Gd$_{0.24}$Fe$_{0.76}$ && 3.2$\degree$ &     35.1$\degree$&    & 0.6$\degree$  &      1.7$\degree$& &0.6$\degree$&  13.6$\degree$& & 0.4$\degree$& 2.2$\degree$     \\ 
Gd$_{0.50}$Fe$_{0.50}$&&    0.6$\degree$&  35.6$\degree$&  &    0.6$\degree$&      4.3$\degree$& &  0.4$\degree$&  10.3$\degree$& &  0.3$\degree$&   1.4$\degree$  \\ 
Gd$_{0.76}$Fe$_{0.24}$ &&      3.2$\degree$&  35.8$\degree$& &  1.2$\degree$&  5.7$\degree$& & 1.2$\degree$&  17.2$\degree$& &0.8$\degree$& 3.2$\degree$    \\ 
\hline
\end{tabular}\label{table3} 
\end{table}

\begin{figure}[ht]
\begin{center}
\includegraphics[angle=0,scale=0.31]{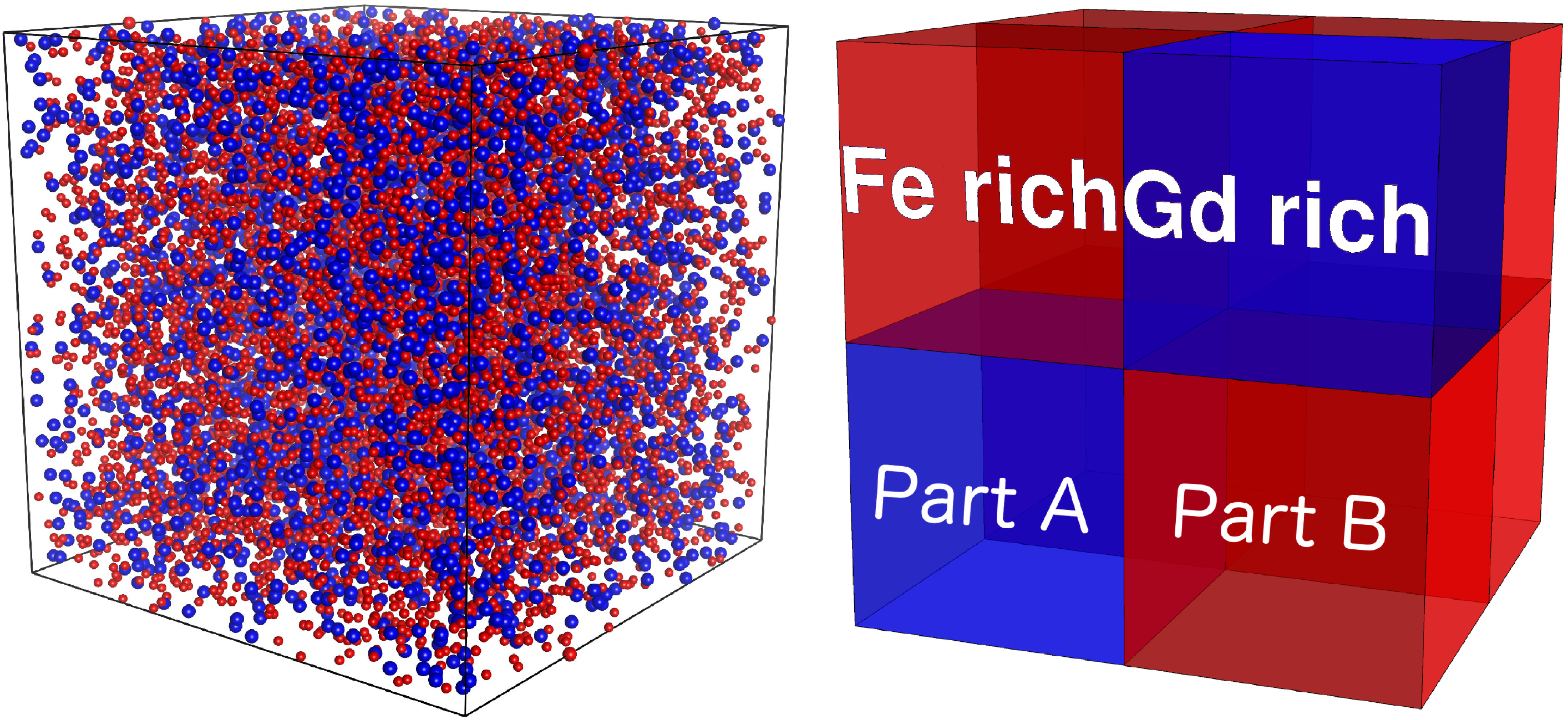}
\caption{\label{figure5} (Color online) Schematic figure showing the inhomogeneous samples of Gd$_{0.24}$Fe$_{0.76}$ plotted in the right side while in the the left side is shown a detailed figure of the amorphous structure with Fe and Gd atoms in red and blue, respectively.  Above, Gd rich regions are called Part A and Fe rich regions are called Part B.}
\end{center}
\end{figure}

\begin{figure}[ht]
\begin{center}
\includegraphics[scale=0.42] {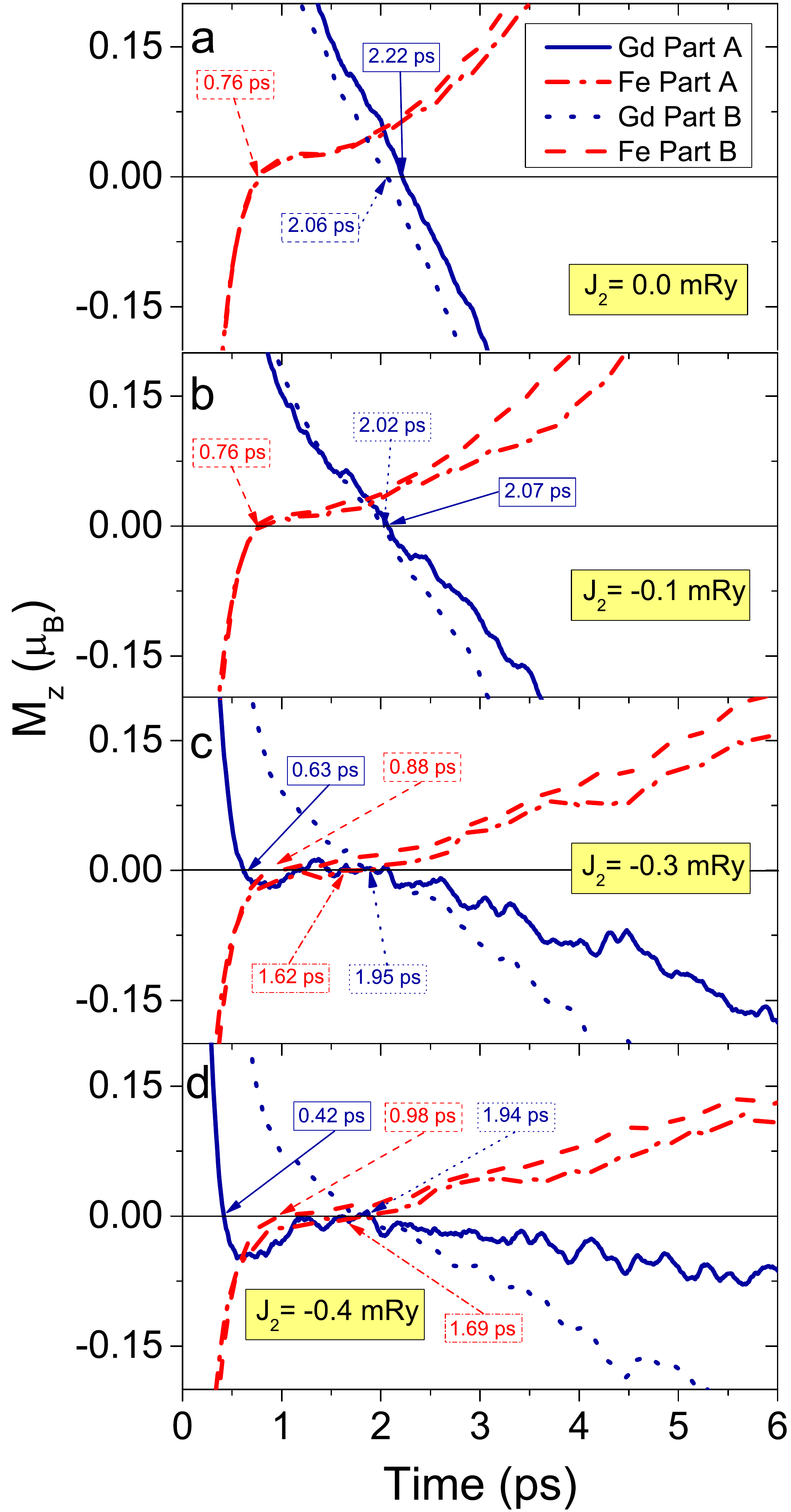}
\caption{\label{figure6}(Color online) Magnetization profile (M$_z$) for Gd (blue) and Fe (red) sublattices for inhomogeneous samples. The magnetization of Gd rich-regions is shown as solid and dash-dot lines (Part A, Gd$_{0.30}$Fe$_{0.70}$) while Gd poor-regions are represented by dash and dot lines (Part B, Gd$_{0.22}$Fe$_{0.78}$). The strength of the second-nearest neighbour exchange parameter (${\it J_2}$) for Gd rich-regions is different for any of the panels (a, b, c and d) outlined in the figure. For Gd poor-regions the {\it J$_2$} values are considered as 0 mRy.}
\end{center}
\end{figure}

The results from the simulation cells with inhomogeneous concentration are shown in Fig.~\ref{figure6}. The figure shows results for four different  degrees of non-collinearity among the Gd atoms in the initial configuration before the heat-pulse enters the spin system. The figure also shows for each panel the sublattice magnetization of the Gd rich-regions  and Gd poor-regions.  Several things may be noted from this figure, where the most important fact is that the experimental results of Ref.~[\onlinecite{graves}] are reproduced in these simulations, if a non-collinear configuration of Gd atomic spins are considered within Gd rich-regions. Figure~\ref{figure6} shows that for a small degree of non-collinear moments of the initial configuration (panels a and b in Fig.~\ref{figure6}), results in a dynamical behavior that is similar to the homogenous results shown in Fig.~\ref{figure3}. Nonetheless, the data shown in Fig.~\ref{figure3} can be considered qualitatively similar to that in Fig.~\ref{figure6} when ${\it J_2}$ is zero, as shown by the fact that there is all-thermal switching and that the Fe sublattice demagnetizes before the Gd sublattice. Moreover, increasing the degree of non-collinearity of the initial configuration causes the Gd moment to demagnetize faster, as show in Fig.~\ref{figure6}, and for sufficiently large values of ${\it J_2}$ it demagnetizes faster than the Fe sublattice. This finding is in agreement with the observation reported in Ref.~[\onlinecite{graves}]. 

\begin{figure}[ht]
	\begin{center}
		\includegraphics[scale=0.36]{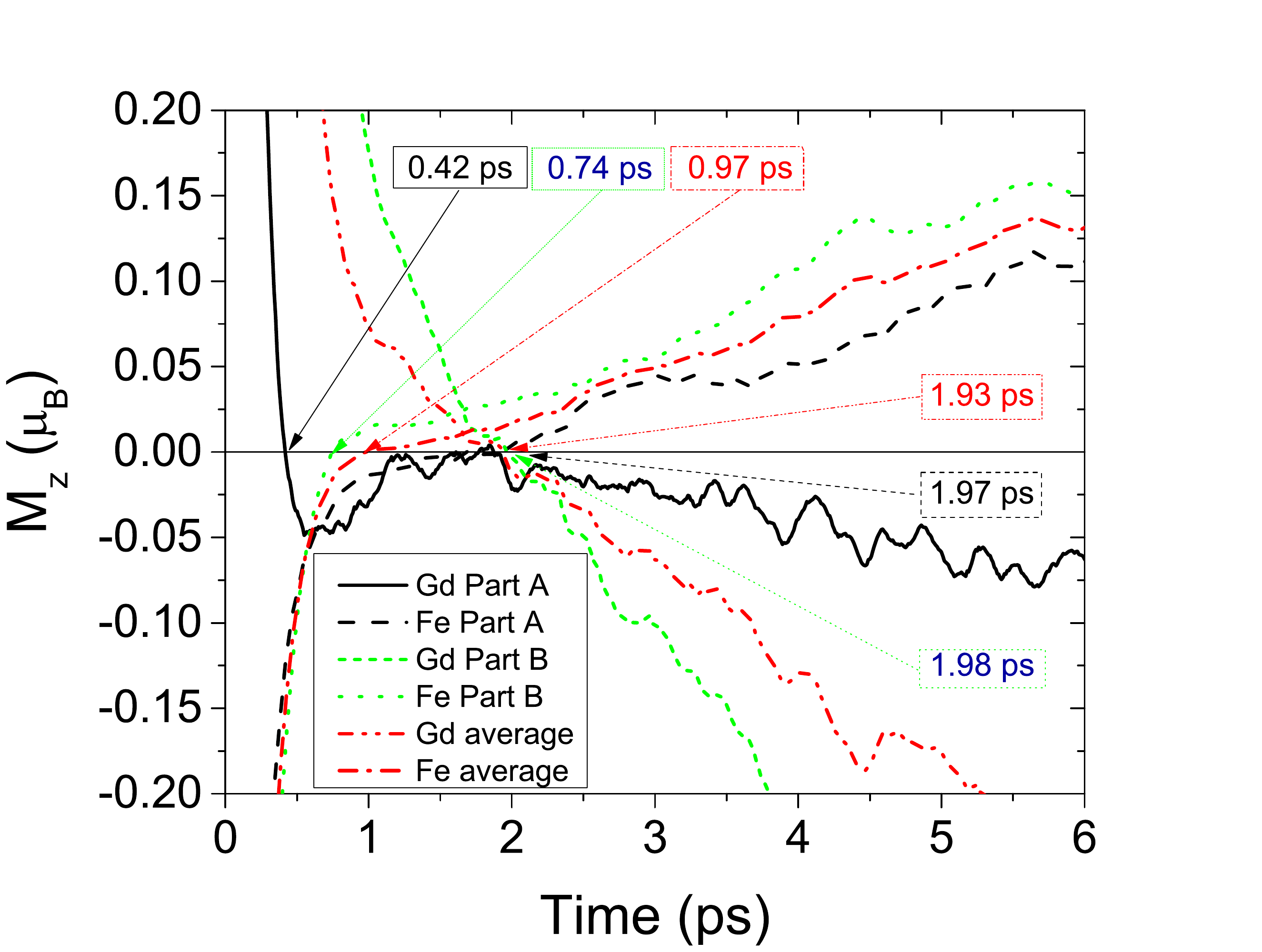}
		\caption{\label{figure7}(Color online) Magnetization profile (M$_z$) for Gd and Fe sublattices for inhomogeneous concentration profiles in a sample with average concentration Gd$_{0.24}$Fe$_{0.76}$. The magnetization of the Gd rich regions (Part A), Gd poor regions (Part B) and sample average is shown in black, green and red lines, respectively.  The strength of the next nearest neighbour exchange parameter (${\it J_2}$) for the Gd sublattice was -0.4 in this simulation. The figure show also similarities with experimental data reported in Fig.~3c of Ref.~[\onlinecite{graves}].}
	\end{center}
\end{figure}

In order to further analyze the results of the heterogenous sample, we show in Fig.~\ref{figure7} the magnetization dynamics of the Gd and Fe sublattices, from the different regions of the sample, i.e. the Gd poor (Fe rich), Gd rich (Fe poor) and regions with an average concentration. We can observe that the Gd sublattice switches faster (0.42 ps) than the Fe one (1.97 ps) in the Gd-rich regions while for the Fe-rich regions the process is reversed. As commented above, this finding was already observed experimentally in Ref.~{\cite{graves}} for GdFeCo. The explanation proposed by the authors relies on the assumption that there are spin currents which transfer torque towards the Gd spins in the enriched Gd nanoregions. The theory put forth here does not involve explicitly a spin current mechanism. Instead, we propose an alternative explanation based on a mechanism driven by a combination of the Dzyaloshiskii-Moriya interaction and exchange frustration that produces non-collinearity of the Gd atoms belonging to the Gd-rich nanoregion. Using this model, the atomistic spin-dynamics simulations show that a non-collinear configuration of spins before a heat-pulse enters the system, explains the faster switching of Gd moments in the Gd rich region. We note that the assumption of non-collinear moments agrees with the non-collinearity predicted by DFT results for the three stoichiometries shown in Fig.~\ref{figure1} (see Appendix~\ref{appendixa}). It is important to emphasize here the role played by the inhomogeneity of the sample. Thus, as shown in Fig.~\ref{figure7}, only the Gd sublattice in the Gd-rich region switches faster than Fe spins while for the sample average, the change of the magnetization occurs first for the Fe sublattice. In order to measure, detect and use that property for technological applications, the experimental techniques are required to have at least a nanometer spatial resolution or lower, as for example, measuring nanometer-femtosecond spin scattering dynamics using X-ray lasers~\cite{graves}.

\begin{figure}[ht]
	\begin{center}
		\includegraphics[scale=0.35]{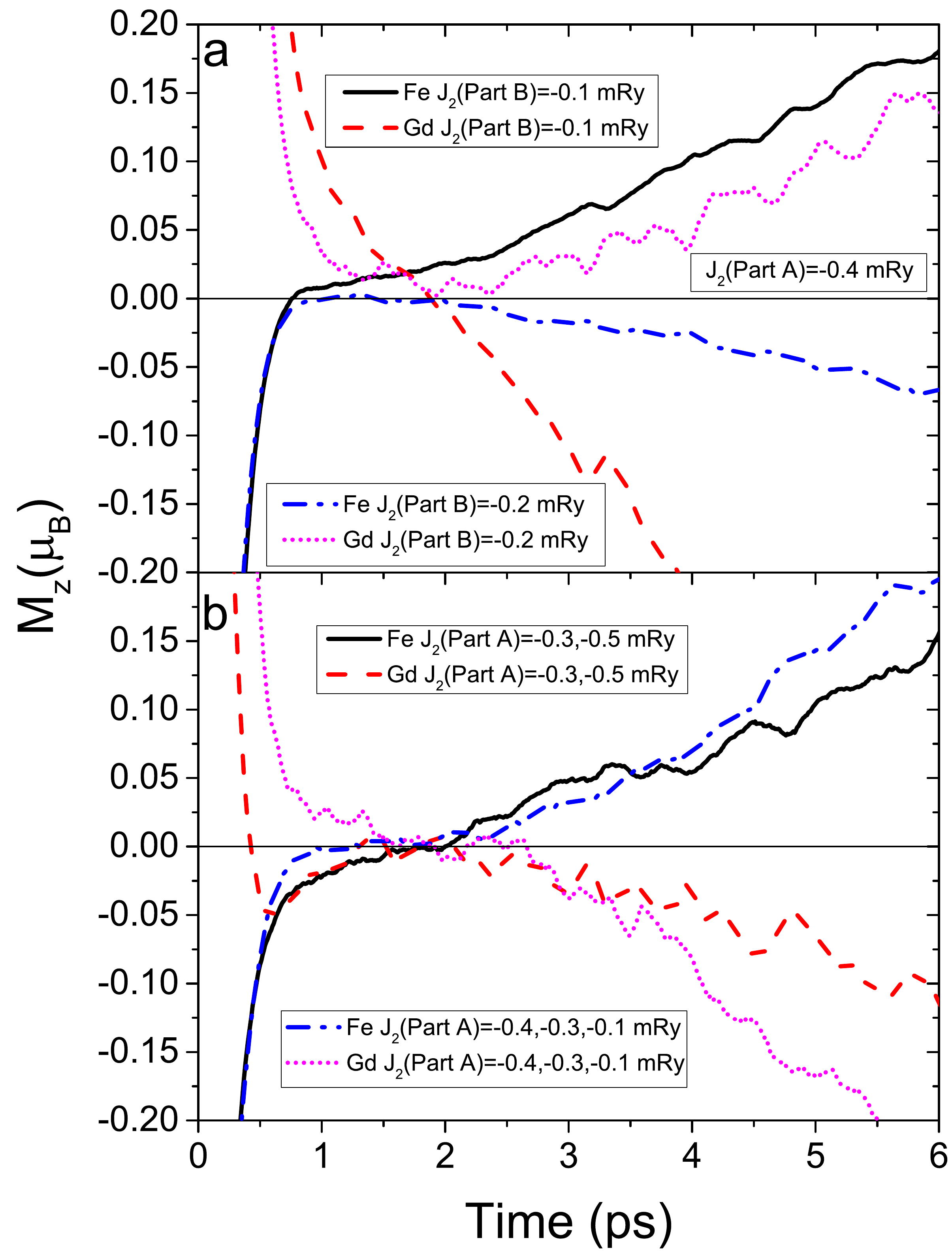}
		\caption{\label{figure8}(Color online) Magnetization profile (M$_z$) for Gd and Fe sublattices for inhomogeneous concentration profiles in a sample with average concentration Gd$_{0.24}$Fe$_{0.76}$. a) The SNN exchange interaction (${\it J_2}$) in the Gd-poor region (Part B) has been chosen to be -0.1 and -0.2 mRy while in the Gd-rich region (part A) ${\it J_2}$=-0.4 mRy. b) Two sets of ${\it J_2}$ parameters distributed randomly over the Gd-rich region. The values of ${\it J_2}$ parameters are listed in the insets of the figure.}
	\end{center}
\end{figure}

Ultimately,  and based on the DFT data collected in Table~\ref{table3}, we observed that the spin-orbit effects contributes to both sublattices, but are more important in Gd sublattices. Moreover, the degree of non-collinearity is not evenly distributed over the Gd and Fe atoms as shown in Table~\ref{table3} by the substantial difference between the average and maximal angle deviation of the atomic magnetic moment. Consequently, the distribution of the  DM vectors or SNN exchange interactions is inhomogeneous within the Gd-rich and Fe-rich regions. In order to mimic and study the effects of the aforementioned inhomogeneous distribution, we performed several ASD simulations with different sets of SNN exchange interactions e.g. using values of -0.3 mRy and -0.5 mRy for the Gd sublattice distributed randomly in the Gd rich region. Interestingly, we observe that an inhomogeneous distribution of the non-collinearity can favour a faster switching of Gd sublattice if, and only if, the four exchange interactions are above a specific threshold (in this set of calculations the threshold was -0.3 mRy, as shown for two sets of parameters in Fig.~\ref{figure8}b). Consequently, an inhomogeneous distribution of non-collinear magnetic moments can also cause Gd to switch faster than Fe only above a minimum value of the degree of non-collinearity. In line with the results described in this section, we also studied the influence of the non-collinearity in the Gd-poor region for Gd sublattice and also for Fe sublattice in the Gd-rich region. We observe in both cases that the increase of the degree of non-collinearity in this situation completely eliminate the switching behavior. We show as an example in Fig.~\ref{figure8}a the magnetization profile of the Gd sublattice in the Gd-poor region for two values of the SNN exchange interaction while for the Gd-rich region the SNN exchange interaction is kept constant as -0.4 mRy. The results show that for ${\it J_2}$=-0.2 mRy, the switching behavior is suppressed for both Fe and Gd sublattices, while if J2 = -0.1 mRy the all thermal switching occurs. The conclusion of all these simulations show that all-thermal switching is determined by delicate details in the concentration profile and the exchange interactions of different regions of the sample. 

\section{Conclusions}

The study of amorphous Gd-Fe alloys represent an outstanding theoretical challenge because of the lack of crystalline periodicity and intrinsic sample inhomogeneities. The intricate structural properties clearly determine the magnetic ones, as usually is the case, and consequently, the magnetization dynamics. We have here been able to address the very complicated structural, magnetic and dynamical properties of several concentrations of amorphous Gd-Fe alloys by using {\it ab initio} DFT in conjunction with an atomistic spin dynamics approach. With the aim to assess the validity of this multi-scale approach, we compare both structural and magnetic parameters with the experimental results and where a comparison can be made, we find that they are in a very good agreement with observed data. In particular, the {\it T$_C$} and {\it T$_M$} predicted by our simulations compare quite well with the experiment and we are able to explain the increase of the compensation temperature with increasing Gd concentration. The explanation mainly resides in the competition of magnetic sublattices with antiparallel coupling. 

Among the most conspicuous results obtained here, we lay emphasis on the crucial role played by the degree of homogeneity and non-collinearity of atomic moments in the Gd-Fe alloys for the thermally induced magnetization switching driven by a femtosecond laser pulse. For homogeneous samples, the Fe sublattice reverse its magnetization before the Gd sublattice for a Gd concentration ranging from 21 at.\% to 30 at.\%. We observe all-thermal switching irrespective of whether the initial temperature was above or below {\it T$_M$}, which is in clear disagreement with previous reported results in literature~\cite{medapalli}. In that regard, the mechanism proposed for all-thermal switching put forward in Ref.~\cite{medapalli} seems to break down for amorphous materials and makes these systems more versatile for spintronic applications since they are less sensitive to the applied initial temperatures. However, for inhomogeneous samples, we found the opposite behavior with respect to homogeneous case, i.e. the Gd sublattice reaches zero magnetization faster than Fe sublattice, at least for the regions with higher Gd concentration. Here, we propose a mechanism based on the influence of Dzyaloshinskii-Moriya interaction and the exchange frustration that we model by considering a second-neighbour exchange interaction between Gd atoms in the Gd-rich regions. The microscopic origin of the antisymmetric Dzyaloshinskii-Moriya interaction is in general known to be coupled to spin-orbit effects and the absence of inversion symmetry, that clearly is present in the amorphous Gd-Fe samples. The influence of the damping parameter was also considered in this work and we observe that this parameter plays a crucial role when dealing with ultrafast switching experiments. Thus, the amorphous Gd-Fe sample with values of $\alpha$ lower or in the vicinity of 0.02 undergo a switching process while for higher values of the damping, the switching mechanism is totally absent. Furthermore, our results point out that a microscopic mechanism for all-thermal switching does not need to involve spin current effects.
 
\begin{acknowledgements}
We gratefully acknowledge financial support from the Swedish Research Council (VR). O.E. is in addition grateful 
to the ERC (project 247062 - ASD) and the KAW foundation for support. J.H.M. acknowledges funding from the Nederlandse Organisatie voor Wetenschappelijk onderzoek (NWO) by a Rubicon Grant. Support from eSSENCE, Stichting voor Fundamenteel Onderzoek der Materie (FOM), De Nederlandse Organisatie voor Wetenschappelijk Onderzoek(NWO), the European Union via ERC Grant agreements No. 257280 (Femtomagnetism), No. 339813 (EXCHANGE), No. 338957 (Femto/Nano) and EC FP7 No. 281043 (FEMTOSPIN) is acknowledged. We also acknowledge Swedish National Infrastructure for Computing (SNIC) for the allocation of time in high performance supercomputers.
\end{acknowledgements}

\appendix
\section{ELECTRONIC PROPERTIES}\label{appendixa}

\begin{figure}[ht]
	\begin{center}
	\includegraphics[scale=0.47]{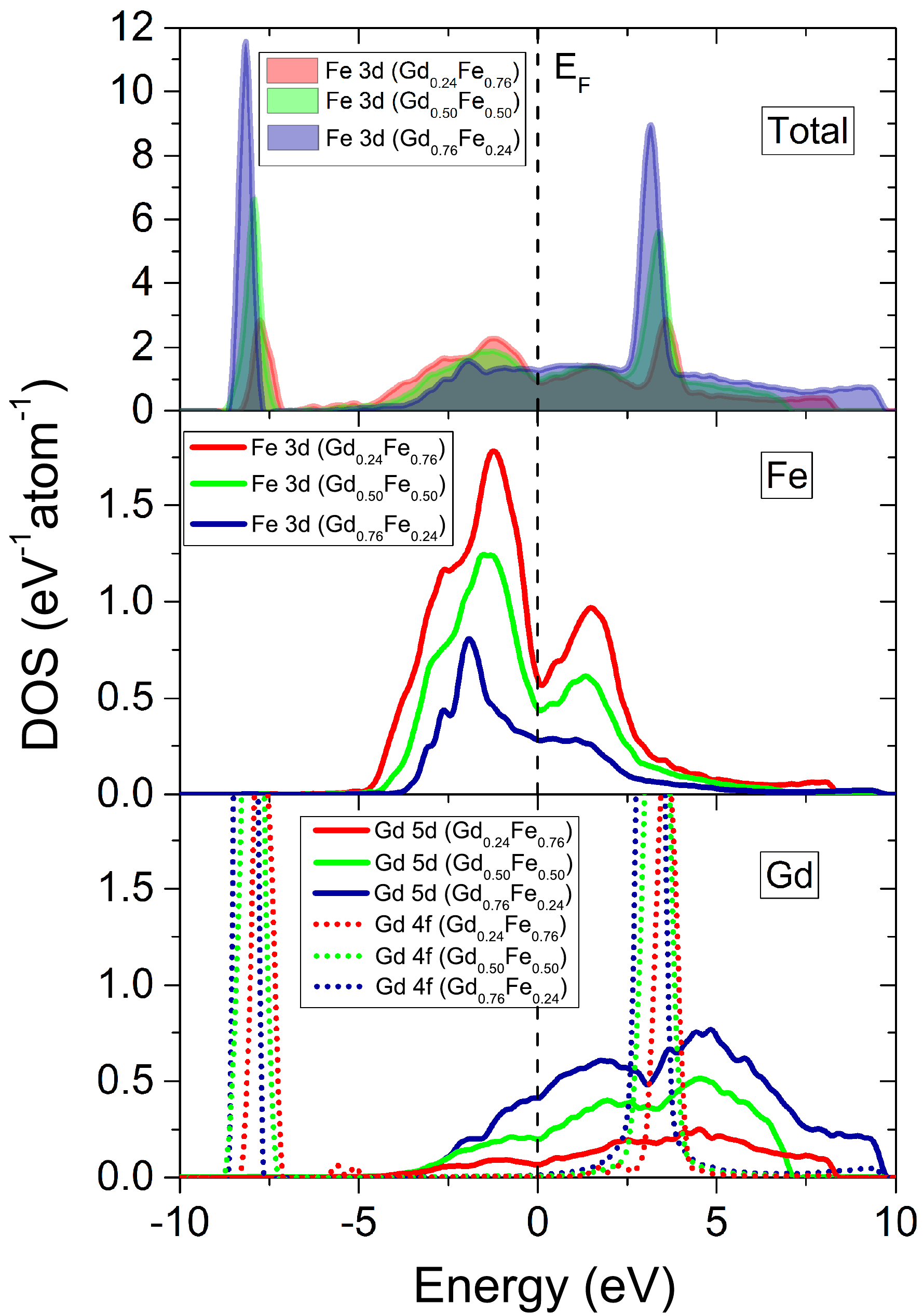}
		\caption{\label{figure9} (Color online) Total and orbital projected  
			densities of states calculated for theoretical Gd$_{x}$Fe$_{1-x}$ structure ($x = 0.24, 0.50, 0.76$). 
			The Fermi level is represented by a dashed vertical line.}
	\end{center}
\end{figure}

Total and partial densities of states (DOS) of the Gd-Fe system are shown in Fig.~\ref{figure9}. 
The metallic character of amorphous Gd-Fe is due to Fe $3d$ and Gd $5d$ states, which contribute
to the DOS at the Fermi level. The Gd $4f$ states in the LDA+U treatment are localized in 
a narrow energy interval around 8 eV below the Fermi level. This is in a rather good agreement with the 
binding energy of the occupied $4f$ states obtained in XPS measurements (9.4 eV)~\cite{guntherodt}. 
Although the LDA+U treatment has been criticized for rare-earths in general~\cite{larspeters}, 
it is shown from our electronic structure calculations to be sufficiently 
accurate for the purposes of the present study. 

Since the electronic structure calculations performed in this work have non-collinear spin densities,
the spin-up or spin-down band picture is no more applicable in our treatment.  
However, the main features of the DOS, i.e. the shape of Gd $4f$ and Fe $3d$ states 
are similar to those calculated for ferrimagnetic amorphous Gd$_{0.33}$Fe$_{0.67}$ alloys 
within DFT+LDA theory \cite{tanaka}. For both approaches, the center of mass of the Fe $3d$ states 
is located below the Fermi level, while the center of mass of Gd $5d$ states is located above 
the Fermi level. With the increase of Gd concentration, the Fe-Fe bond distance 
becomes slightly shorter, while the Fe-Fe coordination number reduces almost 4 times from 7.6 to 2.0. 
This leads to a narrower and less intense Fe $4d$ states in Gd-rich alloy. We can observe, in the 
middle and lower panels of Fig.~\ref{figure9}, that around the Fermi level there is a strong coupling between
Fe $3d$ and Gd $5d$ states. These orbitals are responsible for the strong AFM coupling between Fe and Gd atoms. 

The calculated average magnetic moment is 7.40~$\mu_B$ for Gd and 2.38~$\mu_B$ for Fe atoms. This
is in line with previous experimental data at $T$=4.2 K for amorphous Gd$_{0.33}$Fe$_{0.67}$ 
ferrimagnetic alloy \cite{buschow} and also with theoretical values obtained for Gd$_{0.33}$Fe$_{0.67}$ \cite{tanaka}. In Table~\ref{table3}, we show the average and maximal angles between the z-axis and the magnetic moments on Gd and Fe atoms to estimate the degree of non-collinearity in the amorphous Gd$_{x}$Fe$_{1-x}$ structures ($x = 0.24, 0.50, 0.76$). The results exhibit two well-defined features, i.e., Gd magnetic moments always display angle deviations bigger than Fe magnetic moments regardless of the presence of the spin-orbit coupling in the calculations, and in addition, spin-orbit effects are on the side of an increase of the average angle deviation. Some of the Gd magnetic moments are much larger than the average value, ranging from 13.6 to 17.2$\degree$ without LS coupling and $\sim35\degree$ with LS coupling. The maximal angle deviation tends to be higher as the Gd concentration increases. Thus, the non-collinearity is enhanced in Gd-rich structure. Note that our first principle results contrast the empirical models used to support experimental studies on amorphous alloys, which always assume the Gd sublattice to be collinear \cite{coey}. On the other hand, similar as in our results the degree of non-collinearity of the Fe sublattice increases with increasing the Gd concentration \cite{danh}. If the spin-orbit coupling is present in amorphous Fe-Gd alloys, as is indicated by the current spin-orbit DFT calculations, and due to the fact that amorphous structures lack inversion symmetry, then these conditions create a suitable environment for the DM interaction to be present in these alloys. Even though in amorphous materials it is not possible to apply straightforwardly the usual symmetry-related arguments encompassed by Moriya rules, however it is feasible to use the rather general formulas such as the ones derived in Refs.~\cite{katsnelson,secchi}.

Bader analysis shows that in amorphous Gd$_{x}$Fe$_{1-x}$, there is a charge transfer from Gd to Fe atoms. With the increase of Gd concentration from 24 at.\% to 76 at.\%, Fe atoms gain more negative charge, simply due to the fact that the probability of finding Gd atoms located in nearest neighbour positions around an Fe atom increases. At the same time, the average valence electron increases by 0.73 and 0.53 for Gd and Fe atoms, respectively.

\section{BINDER CUMULANT}
\label{bindercumulants}

\begin{figure}[ht]
\begin{center}
\includegraphics[scale=0.48]{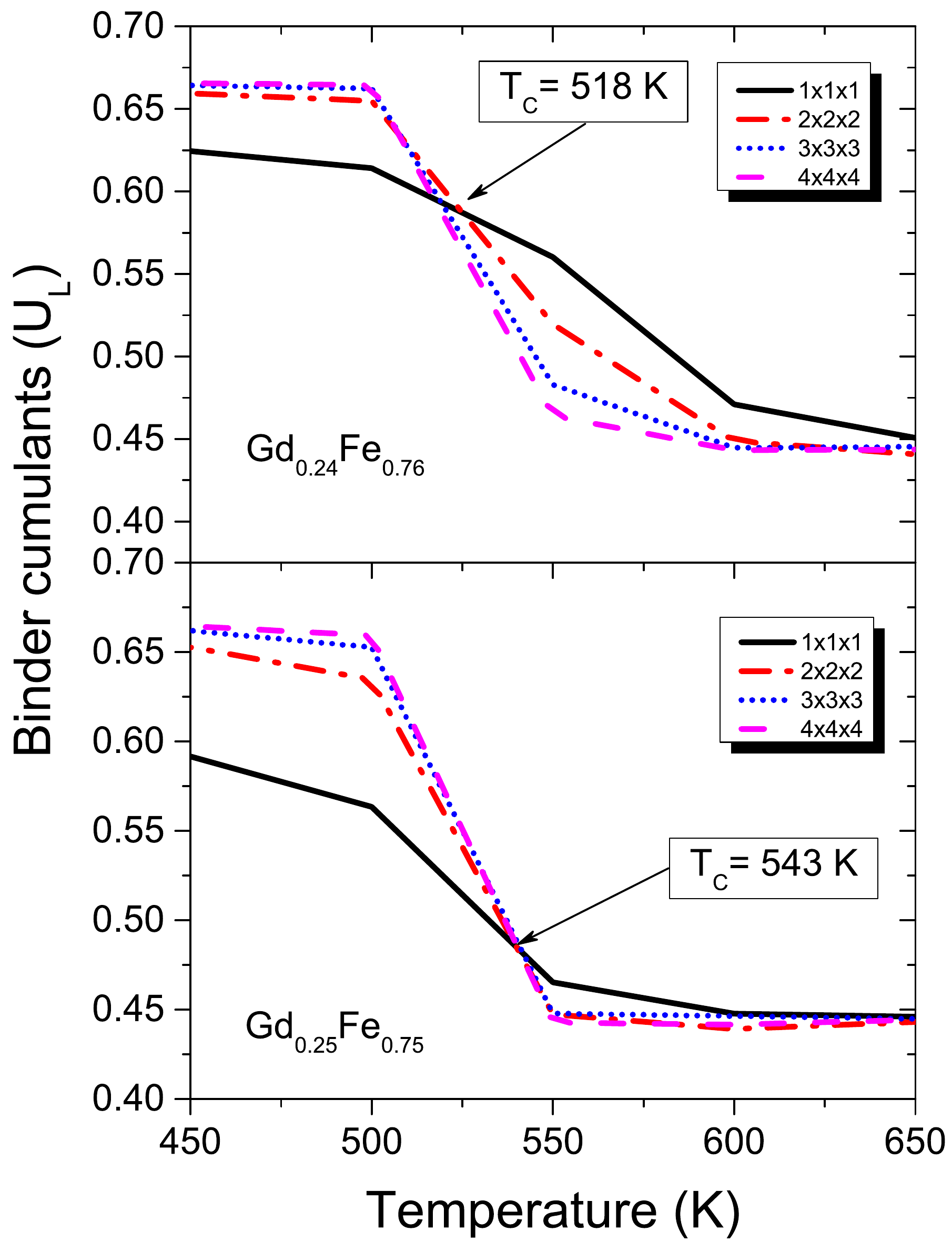} 
\caption{\label{figure10} (Color online) Calculated Binder cumulants for Gd$_{0.24}$Fe$_{0.76}$ and Gd$_{0.25}$Fe$_{0.75}$ samples with 1600 atoms per unit cell. The size of the samples ($x\times y\times z$) was ranged from $x=y=z=1$ to 4 unit cells in steps of 1. The Curie temperature is indicated by the arrows. }
\end{center}
\end{figure}

The fourth order Binder cumulant was introduced in Ref.~[\onlinecite{binder}] in the context of the finite size scaling theory~\cite{fisher}. For magnetic atoms arranged in a lattice of size L, the Binder cumulant is defined by:

\begin{equation}
\label{eq_binder}
U_L=1-\frac{<m^4>_L}{3<m^2>^2_L}
\end{equation}
where $m$ is the order parameter, i.e. the magnetization and $<>$ denotes the statistical average taken over systems at equilibrium and at constant temperature. The Binder cumulant allows to locate the critical point and the critical exponents in a phase transition. Thus, in the thermodynamic limit where the system size of the ferromagnet $L\rightarrow\infty$ and consequently $L$ is bigger than the  correlation length, the Binder cumulant approximates to zero for temperatures higher than {\it T$_C$} while for temperatures lower than {\it T$_C$}, ${\it U_L}\rightarrow\frac{2}{3}$. This property of the cumulant is very useful for obtaining very good estimates of {\it T$_C$} which are not biased by any prerequisites about critical exponents. After performing the atomistic spin dynamics simulations, we have access to the magnetization which is inserted in Eq.~(\ref{eq_binder}) to obtain as a result the Binder cumulant. Then, we plot the cumulant as a function of the temperature for different sample sizes and {\it T$_C$} is estimated from the intersection point of those curves. In Fig.~\ref{figure10} we show, as an example, the Binder cumulants for two Gd concentrations and the estimated {\it T$_C$} for both samples. The Curie temperatures shown in Fig.~\ref{figure2} have been calculated using the procedure described above.

\section{DETAILS OF THE TWO-TEMPERATURE MODEL}
\label{2tmodel}

In order to study the ultrafast demagnetization, the three temperature model (3TM) was introduced by Beaurepaire {\it et al.} in 1996~\cite{beaurepaire}. In the 3TM model, the electron, spin and phonon are thermal reservoirs and coupled to each other by coupling constants. It is difficult to define an electron temperature for the first femto-seconds of a Laser induced pump-probe experiment, but we have here for simplicity used the 3TM model. The analytical expression of the 3TM contains three differential equations and from that equations, the three temperatures {\it T$_{e}$} (electron temperature), {\it T$_{s}$} (spin temperature) and {\it T$_{latt}$} (lattice or phonon temperature) are calculated. In equilibrium the temperature of all three thermodynamic reservoirs is equilibrated, i.e. {\it T$_{e}$}={\it T$_{s}$}={\it T$_{latt}$. In this model we assume that the lattice is an infinitely large thermal reservoir with constant temperature {\it T$_{latt}$}. This assumption seems to be quite valid as it was reported in typical pump-probe experiments \cite{koopmans1}. Moreover, we also assume that the electron reservoir is a thermal reservoir much smaller than the lattice, but still larger than the spin system. The spin temperature is explicitly passed into the stochastic LLG equation while the electron temperature can be expressed in a simple analytical form as, 
\begin{eqnarray}
\label{temperature_electron}
T_{e} & =& T_{0} + (T_{P}-T_{0})\cdot(1- e^{(-t/\tau_{i})})\cdot e^{(-t/\tau_{f})}  \\
 &+& (T_{F}-T_{0})\cdot(1-e^{(-t/\tau_{f})})\nonumber
\end{eqnarray}
Thus, we reduced the 3TM model into a simple exponential function~\cite{mentink, skubic}, that captures the essential physics of the three temperature model.  In Eq.~\ref{temperature_electron},} {\it$T_{0}$} represents the initial temperature of the system, {\it T$_{P}$} is the maximum temperature achieved in the simulation and {\it $T_{F}$} is the final temperature of the system. The function depends on two time parameters, such as initial time $\tau_{i}$ and final time $\tau_{f}$. The initial time describes the rise-time of the temperature to its maximum value and the final time represents the relaxation time of the temperature from the maximum value to the final temperature of the system. Here we used $\tau_{i}=50$ fs and $\tau_{f}=1$ ps. In this model, the electron temperature is used as spin temperature. At each time step, the calculated spin temperature is passed explicitly into the stochastic field of the LLG equation. The values of {\it T$_{P}$}=800 K, 1000 K and 2000 K and {\it $T_0$}={\it T$_{F}$}=300 K are used in the simulations. 

\section{DAMPING EFFECTS}
\label{damping_effect}

One of the main parameters in LLG equation is the Gilbert damping, $\alpha$, which is mainly responsible for bringing the system into an equilibrium state. It was experimentally observed that the damping constant $\alpha$ significantly depends on the Gd content and it becomes large near to the compensation temperature of the samples~\cite{kirilyuk}. Though the g-factors of Gd and Fe sublattice magnetic moments in our samples are slightly different, we used same g-factor for both sublattices, and in our initial simulations, the static damping parameter was also kept equal for both sublattices in the atomistic spin dynamics simulations. With these parameters, the evolution of the magnetization of Gd and Fe moments under the influence of an intense femtosecond laser pulse shows different precession and reproduce experimental observations. 

\begin{figure}[ht]
\begin{center}
\includegraphics[scale=0.49]{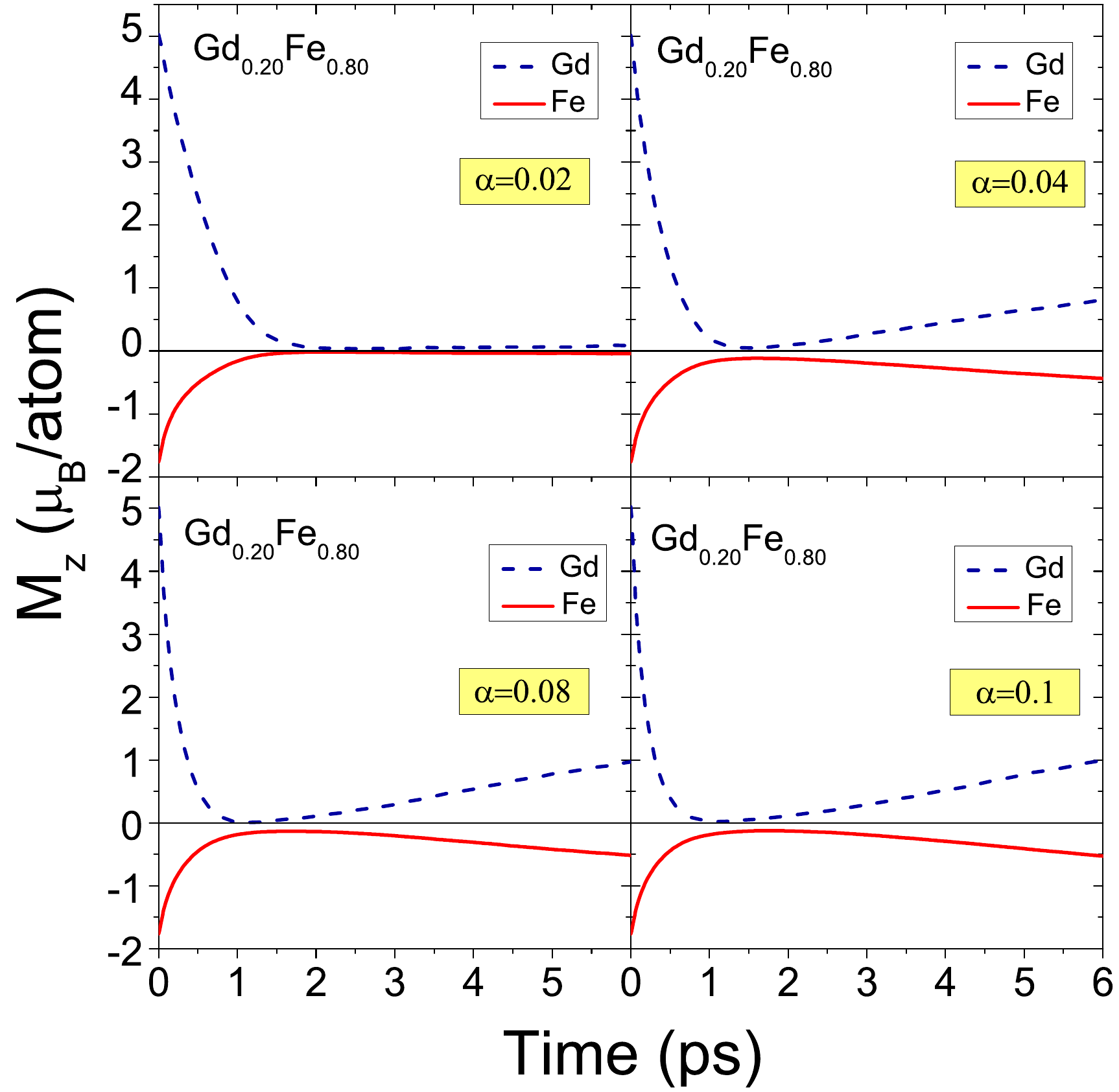}
\end{center}
\caption{(Color online) Time dependence of ultrafast magnetization (M$_z$) in Gd$_{0.20}$Fe$_{0.80}$ for different 
damping parameters ($\alpha = $ 0.02, 0.04, 0.08, 0.1)  with a fixed electron temperature profile.}
\label{figure11}
\end{figure}
\begin{figure}[ht]
\begin{center}
\includegraphics[scale=0.49]{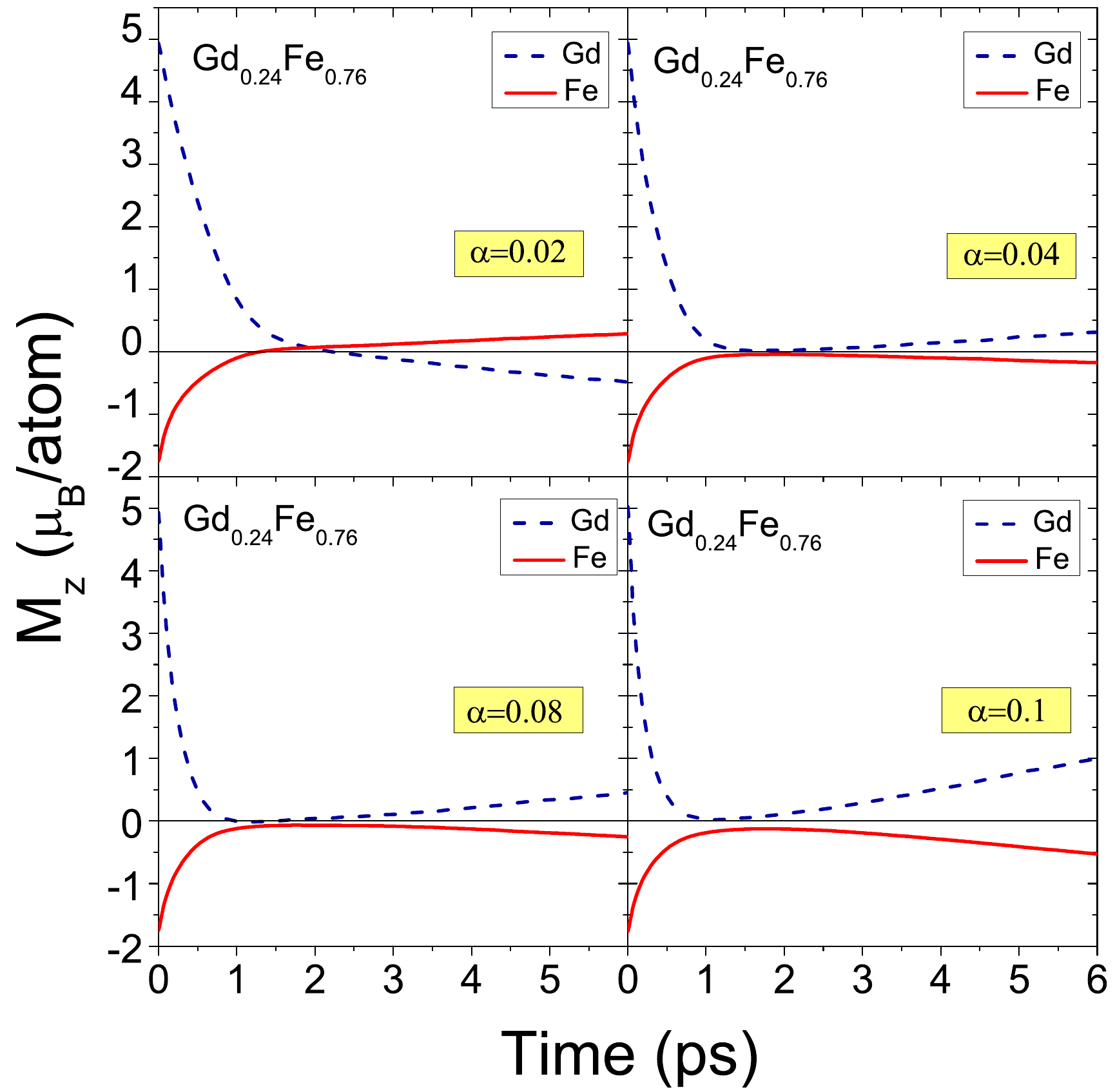}
\end{center}
\caption{(Color online) Time dependence of ultrafast magnetization (M$_z$) in Gd$_{0.24}$Fe$_{0.76}$ for four different 
damping parameters ($\alpha = $ 0.02, 0.04, 0.08, 0.1) with a fixed electron temperature profile.}
\label{figure12}
\end{figure}
\begin{figure}[ht]
\begin{center}
\includegraphics[scale=0.49]{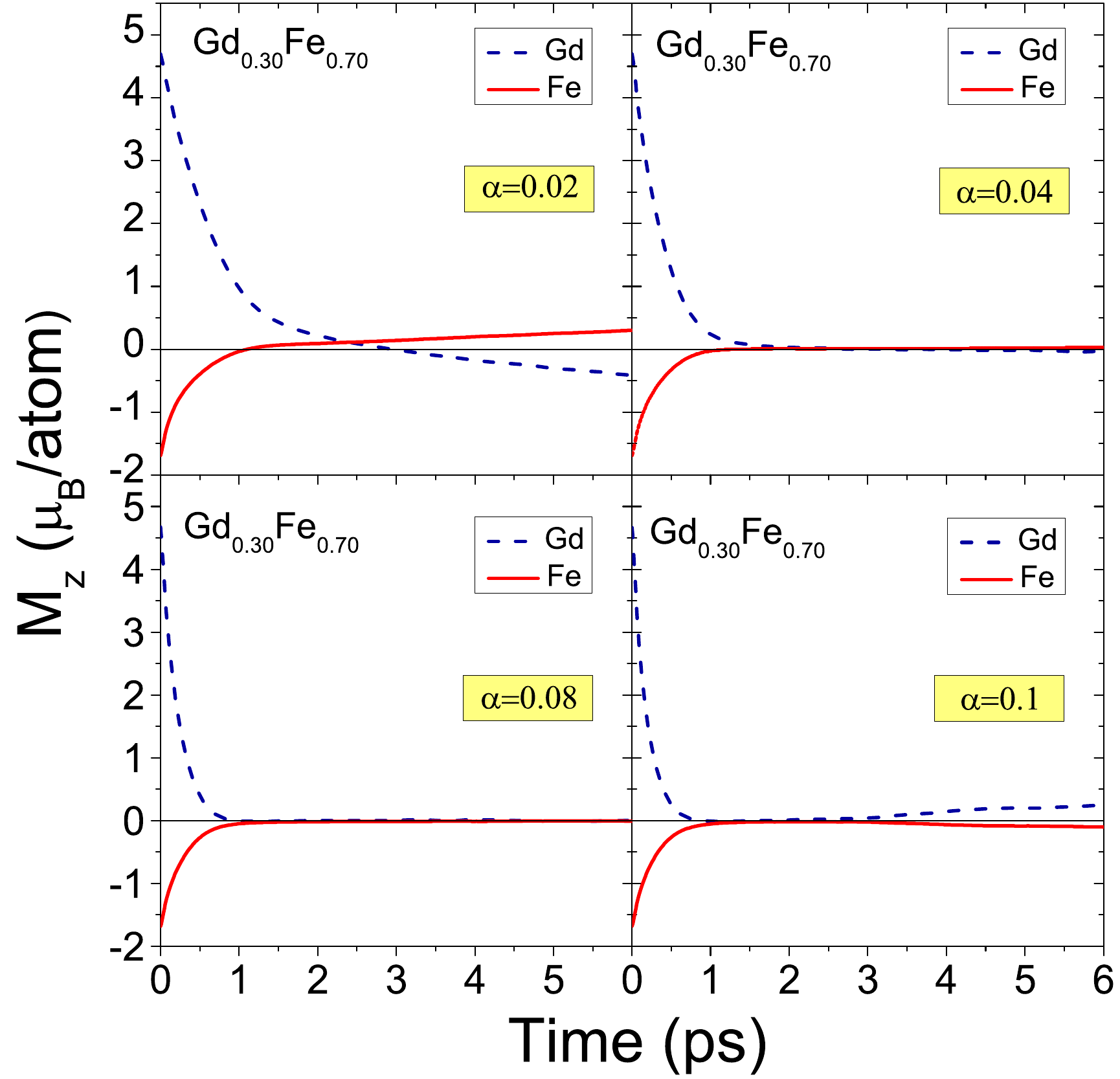}
\end{center}
\caption{(Color online) Time dependence of ultrafast magnetization (M$_z$) in Gd$_{0.30}$Fe$_{0.70}$ for four different 
damping parameters ($\alpha = $ 0.02, 0.04, 0.08, 0.1) with a fixed electron temperature profile.}
\label{figure13}
\end{figure}

Later on, in a second stage of our simulations, we adapted a site-dependent damping parameters in the LLG equation and we performed ultra-fast simulations on  Gd$_{0.20}$Fe$_{0.80}$, Gd$_{0.24}$Fe$_{0.76}$ and Gd$_{0.30}$Fe$_{0.70}$ amorphous alloys. These results are shown in Figs.~\ref{figure11}-\ref{figure13}. By fixing the damping 
parameter of Fe species as $\alpha_{Fe}=0.02$, we changed the damping parameter of Gd ($\alpha_{Gd}$) from 0.02 to 0.1 in steps of 0.02. If $\alpha_{Gd}=0.02$, 
the simulations for the three sample concentrations shown in Figs.~\ref{figure11}-\ref{figure13} predict that the Fe sublattice demagnetize faster than the Gd sublattice. Samples with compensation temperatures, i.e. with a composition of Gd ranging from 21 at.\% to 30 at.\%, shows switching behavior (see Fig.~\ref{figure2}). In Fig.~\ref{figure11}, we found that for Gd$_{0.20}$Fe$_{0.80}$ amorphous sample, there is no ultrafast switching. This result clearly shows that the compensation point is a very important parameter in the spin dynamics of Gd-Fe alloy. For higher concentrations of Gd, we observed the switching behavior, as shown in Figs.~\ref{figure12}-\ref{figure13}. In our calculations, we found the transition metal demagnetize faster than the rare-earth element and forms a ferromagnetic-like state for a short period of time due to the AFM interaction between Gd and Fe atoms. This was already explained in Ref.~[\onlinecite{radu}]. The idea is that the AFM coupling between Gd and Fe atoms favors the spin flipping of Fe atoms when Gd atoms are becoming reversed. Thus, the process promotes an increase of the net Fe magnetization parallel to the remaining Gd magnetization.

In the case that $\alpha_{Gd} > 0.02$, we observed that the Gd sublattice moves towards 
sub-picosecond times but never becomes FM to Fe sublattice, as shown in Figs.~\ref{figure11}-\ref{figure13}. The main message of these results is that the damping is a very crucial parameter in ultrafast switching process.

\section{GENERATING INHOMOGENEOUS SAMPLES}
\label{inhomogeneous}
\begin{figure}[H]
	\begin{center}
		\includegraphics[scale=0.38]{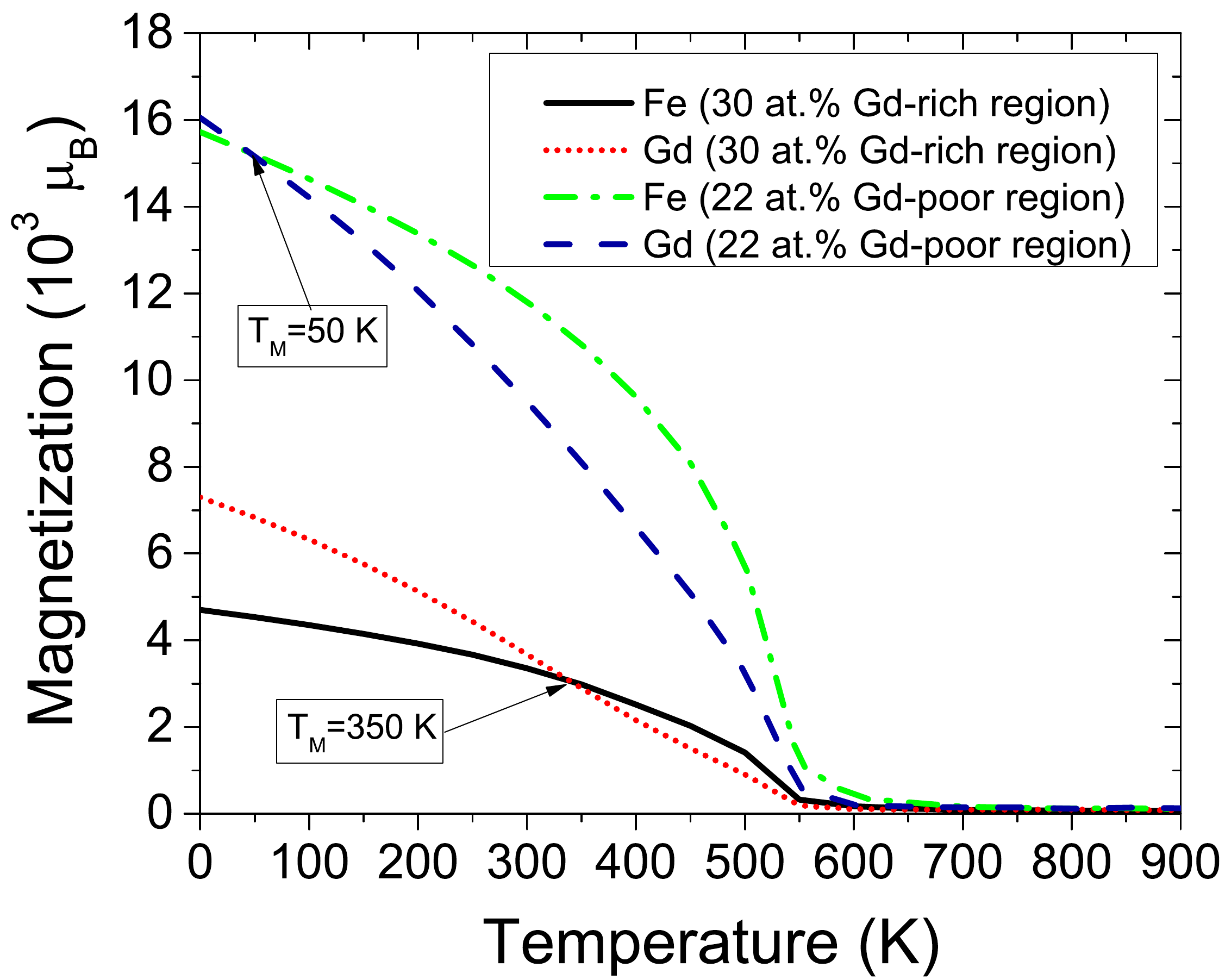}
		\caption{\label{figure14}(Color online) Fe and Gd sublattice magnetizations for Fe and Gd rich-regions plotted against the temperature. The sublattice magnetization is also shown for regions with higher (30 at.\%, solid and dot lines) and lower (22 at.\%, dash and dash-dot lines) Gd concentration. The figure is plotted for {\it J$_2$}=0 mRy.}
	\end{center}
\end{figure}

The inhomogenous sample  was constructed from the homogeneous unit cell of Gd$_{0.24}$Fe$_{0.76}$ by repeating the unit cell twice in  x, y and z direction (12800 atoms), as shown in the right panel of Fig.~\ref{figure5}. Then, a cube is selected randomly and we reduced the percentage of Fe by 6 at.\%. Thus, the cube turns out to be Gd rich region and in the same way the Fe rich regions were created. The new supercell consists of an inhomogeneous environment and it mimics the original samples of GdFeCo experimental results. After that, we study the compensation temperatures for rich and poor areas of Gd. The obtained compensation temperatures are similar to homogenous unit cells. As shown in Fig.~\ref{figure14}, we obtained a compensation temperature of about 50 K and 350 K for Gd poor and Gd rich areas, respectively. The results shown in Appendix~\ref{damping_effect} clearly pinpoint an impossibility of Gd sublattice to switch first than the Fe one. Thus, in Sec.~\ref{inhomogeneous_section},  we incorporate non-collinearity in the sample by introducing an extra AFM exchange value to Gd sublattice in the Gd rich areas. Such type of effects are observed in the experimental samples. The origin of those effects can mainly reside in the concentration of Gd-Fe amorphous samples, which modify their structure-sensitive properties, such as the magnetic ones, compensation temperatures between the Gd rich and poor regions and also sperrimagnetism found in the Fe sublattice of Gd-Fe amorphous alloys for higher Gd-concentration while the collinear structure may be expected to exist for Gd-poor alloys~\cite{danh}.  

\end{document}